\documentclass[11pt,a4paper]{article}
\usepackage[utf8]{inputenc}
\usepackage{amsmath, amssymb, amsfonts, bm}
\usepackage{booktabs} 
\usepackage{graphicx} 

\usepackage{xcolor}   
\usepackage{geometry}
\usepackage{hyperref}
\usepackage{comment}
\usepackage{xurl}
\usepackage[section]{placeins}

\title{A High-Order Nodal Galerkin Formulation for the M\"{u}ller Equation: Bypassing Divergence Conformity via Kernel Cancellation}

\author{Yao~Luo
\thanks{Y. Luo is with the School of Electrical Engineering and Automation, Wuhan University, Wuhan, 430072, China. He is currently a Visiting Scholar with the Dipartimento di Elettronica, Informazione e Bioingegneria (DEIB), Politecnico di Milano, 20133 Milan, Italy (e-mail: sturmjungling@gmail.com).}
}

\date{}

\begin{document}

\maketitle

\begin{abstract}
The M\"{u}ller boundary integral equation for penetrable electromagnetic scattering is conventionally discretized using divergence-conforming basis functions, a restriction inherited from the PMCHWT framework. This paper demonstrates that this constraint can be bypassed. The double-gradient operator in the M\"uller formulation acts on the kernel difference $\varphi_a - \varphi_i$, so that the $\mathcal{O}(R^{-3})$ hypersingularity cancels identically, reducing the operators to weakly singular $\mathcal{O}(R^{-1})$ kernels. Exploiting this cancellation, we develop a nodal, high-order Galerkin formulation using $\mathrm{P}_2$ isoparametric shape functions on curved manifolds. The surface vector field is constructed via a metric-weighted orthonormal tangent frame. The singular integrals are evaluated by Sauter--Schwab quadrature, and a Morton-ordered Block Jacobi preconditioner is introduced. By capturing the dominant near-field interactions within geometrically clustered diagonal blocks, it yields robust, superlinear GMRES convergence under extreme material and geometric parameters. Validation against semi-analytical EBCM references confirms high-order spatial accuracy and optical-theorem satisfaction to high precision.
\end{abstract}

\section{Introduction}
\label{sec:introduction}

The simulation of electromagnetic scattering by penetrable dielectric and lossy media is a central problem in optics, plasmonics, and
microwave engineering. Boundary integral equations (BIEs) are favored for such open-region problems due to their exact enforcement of the Silver--M\"{u}ller radiation condition~\cite{Silver1949, Mueller1951,
Mueller1957} and their reduction of dimensionality. Among the available formulations, the Poggio-Miller-Chang-Harrington-Wu-Tsai (PMCHWT)~\cite{Poggio1973, Chang1977, Wu1977} and M\"{u}ller BIEs are the two most prominent coupled systems. While the PMCHWT formulation is widely used in commercial and academic solvers, M\"{u}ller BIE~\cite{Mueller1951, Mueller1957} offers immunity to spurious interior resonances and a Fredholm second-kind structure, which guarantees a bounded condition number as the mesh is refined.

Despite these theoretical advantages, the numerical implementation of M\"{u}ller's equation has been constrained by an inherited
methodological paradigm. In standard Method of Moments (MoM) treatments, BIEs involving the Hessian of the scalar Green's function
are presumed to possess $\mathcal{O}(R^{-3})$ hypersingularities. The standard remedy is regularization via integration by parts, which
transfers the spatial derivatives from the kernel onto the trial and test functions. This technique requires divergence-conforming basis
functions, most notably the Rao--Wilton--Glisson (RWG) edge elements~\cite{Rao1982}. As a result, the M\"{u}ller equation is almost exclusively treated in the literature as a parameterized
variant of the PMCHWT system~\cite{Oijala2005, Bogaert2014}, forcing it into the same divergence-conforming framework \cite{Yan2013}.

Imposing divergence conformity, however, introduces computational difficulties when extending solvers to high-order curved manifolds.
Classical curvilinear vector bases~\cite{Graglia1997} maintain normal
flux continuity but embed local metric parameters into the interpolation polynomials. This couples the algebraic basis to the
differential geometry, complicating the isoparametric mapping and imposing additional topological bookkeeping during matrix assembly.
Furthermore, the integration-by-parts regularization may generate auxiliary line integrals that burden the numerical quadrature. Together, these constraints limit the flexibility needed to deploy high-order
geometric representations.

The central motivation of this paper is the recognition that this edge-based treatment is an inherited algorithmic constraint rather than
a necessity of the M\"{u}ller formulation. In M\"{u}ller's original construction, the Hessian operator acts exclusively on the kernel difference $\varphi_a - \varphi_i$. Because the static limits of the exterior and interior kernels are identical, the $\mathcal{O}(R^{-3})$ hypersingularity cancels at the operator level,
and the remaining kernel contributions are at most weakly singular $\mathcal{O}(R^{-1})$. Once this cancellation is isolated, the transfer of derivatives onto the surface-current basis is no longer necessary.

Exploiting this cancellation, we present a high-order nodal Galerkin
formulation for the M\"{u}ller system that abandons
divergence-conforming bases. The equivalent electric and magnetic
surface currents are discretized using high-order isoparametric shape
functions, with the vector orientation governed by an orthonormal tangent frame. The construction of this frame
on general curved manifolds requires a robust nodal normal estimation.
Rather than relying on standard area-weighted normal averaging, which
introduces geometric inconsistencies on distorted
meshes~\cite{Meyer2003}, we generalize Max's vertex-angle weights~\cite{Max1999} to high-order curved surfaces via a covariant metric formulation. The generalized weighting exploits the covariant basis vectors to incorporate curvature information from adjacent curved elements. By tying the vector orientation of the currents to this frame, the differential geometry of the scatterer is decoupled from the algebraic amplitude interpolation. The weakly singular kernels are evaluated through the Sauter--Schwab quadrature (SSQ)~\cite{SauterSchwab2011}, which bypasses analytical extraction and accommodates high-order elements
via standard tensor-product Gauss rules.

While the Fredholm second-kind nature of the M\"{u}ller operator ensures $h$-independent GMRES convergence, extreme material parameters and complex
geometries can still degrade the spectral properties of the discrete system. To maintain rapid convergence under such conditions, the resulting
linear system is solved with GMRES equipped with a Morton-ordered
Block Jacobi (MBJ) preconditioner. The Morton
ordering~\cite{Morton1966, Bader2013} maps three-dimensional spatial
proximity to one-dimensional memory contiguity, concentrating the
dominant near-field couplings along the main diagonal and enabling a
block-local preconditioner to capture them. The complete framework is validated through three benchmarks comprising four test cases: a gold spheroid under oblique illumination, the LSPR (localized surface plasmon resonance) spectrum \cite{Bohren1998, Maier2007} of a silver spheroid, a dielectric spheroid at elevated electrical size, and a non-convex Chebyshev
particle.

This work demonstrates that once M\"{u}ller's original cancellation structure is exploited, high-order nodal elements provide a consistent
and computationally efficient alternative to edge-based integral equation solvers.

\section{M\"uller Integral Formulation and Singular-Kernel Cancellation}
\label{sec:mueller_formulation}

\subsection{Strong Form of the M\"uller System}

This work is based on the boundary integral formulation of
M\"uller \cite{Mueller1951,Mueller1957}, which yields a coupled second-kind
system for penetrable dielectric scatterers. Let $\Gamma$ denote the
material interface separating the exterior medium $(a)$ from the interior
medium $(i)$. The scalar Helmholtz Green's function is written as
\begin{equation}
\varphi_{a,i}(R) = \frac{e^{ik_{a,i}R}}{4\pi R},
\qquad
R = \|\bm{x} - \bm{y}\|,
\label{eq:green_kernel}
\end{equation}
where $\bm{x}$ and $\bm{y}$ are the observation and source points,
respectively, and a time dependence of $e^{-i\omega t}$ is assumed
throughout. The wave number in each medium is
$k_{a,i} = \omega\sqrt{\varepsilon_{a,i}\mu_{a,i}}$, where $\omega$ is the angular frequency and $\mu_{a,i}$ the permeability. The complex permittivity $\varepsilon_{a,i}$ is defined as
\begin{equation}
\varepsilon_{a,i} = \varepsilon'_{a,i} + i\frac{\sigma_{a,i}}{\omega},
\end{equation}
with $\varepsilon'_{a,i}$ the real permittivity and $\sigma_{a,i}$ the
electrical conductivity.

Following the modern equivalence-principle convention, the equivalent
electric and magnetic surface currents are defined as
\begin{equation}
\bm{j} = \bm{n} \times \bm{H},
\qquad
\bm{j}_m = -\bm{n} \times \bm{E}.
\label{eq:current_definitions}
\end{equation}
M\"uller's original derivation adopted definitions that differ from these
by a global sign reversal
($\bm{j}_{\text{M\"uller}} = -\bm{n} \times \bm{H}$ and
$\bm{j}_{m,\text{M\"uller}} = \bm{n} \times \bm{E}$); however, because
both definitions change sign simultaneously, the factor cancels from both
sides of the BIEs, leaving them in their exact
original form.

Under the present convention, the M\"uller BIEs for the equivalent electric current $\bm{j}$ and magnetic current $\bm{j}_m$
read
\begin{equation}
\begin{aligned}
&\frac{\varepsilon_i+\varepsilon_a}{2}\,\bm{j}_m(\bm{x})
=
\varepsilon_a\,\bm{j}_m^{\mathrm{inc}}(\bm{x})
-\int_\Gamma
\bm{n}(\bm{x})\times
\Big[
\bm{j}_m(\bm{y})\times \nabla_{\bm{x}}\bigl(\varepsilon_a\varphi_a-\varepsilon_i\varphi_i\bigr)
\Big]
\,d\Gamma_{\bm{y}}\\
&\quad
-\frac{i}{\omega}\int_\Gamma
\bigg\{
\big[\bm{n}(\bm{x})\times \bm{j}(\bm{y})\big]
\bigl(k_a^2\varphi_a-k_i^2\varphi_i\bigr)
+\bm{n}(\bm{x})\times
\big(\bm{j}(\bm{y})\cdot \nabla\big)\nabla\bigl(\varphi_a-\varphi_i\bigr)
\bigg\}
\,d\Gamma_{\bm{y}} ,
\end{aligned}
\label{eq:mueller_electric}
\end{equation}

and

\begin{equation}
\begin{aligned}
&\frac{\mu_i+\mu_a}{2}\,\bm{j}(\bm{x})
=
\mu_a\,\bm{j}^{\mathrm{inc}}(\bm{x})
-\int_\Gamma
\bm{n}(\bm{x})\times
\Big[
\bm{j}(\bm{y})\times \nabla_{\bm{x}}\bigl(\mu_a\varphi_a-\mu_i\varphi_i\bigr)
\Big]
\,d\Gamma_{\bm{y}} \\
&\quad
+\frac{i}{\omega}\int_\Gamma
\bigg\{
\big[\bm{n}(\bm{x})\times \bm{j}_m(\bm{y})\big]
\bigl(k_a^2\varphi_a-k_i^2\varphi_i\bigr)
+\bm{n}(\bm{x})\times
\big(\bm{j}_m(\bm{y})\cdot \nabla\big)\nabla\bigl(\varphi_a-\varphi_i\bigr)
\bigg\}
\,d\Gamma_{\bm{y}} .
\end{aligned}
\label{eq:mueller_magnetic}
\end{equation}
Here $\bm n(\bm{x})$ is the outward unit normal at the observation point
$\bm{x}$, directed from the interior into the exterior medium. The
operator $\nabla_{\bm{x}}$ acts on the observation coordinate, while $\nabla$
acts on the source coordinate. The excitation terms $\bm j_m^{\mathrm{inc}}$ and
$\bm j^{\mathrm{inc}}$ are determined by the tangential components of the incident
fields. Several typographical sign errors present in M\"uller's original
manuscripts \cite{Mueller1951,Mueller1957} have been corrected here.

Unlike conventional formulations that regularize hypersingular kernels through integration by parts in the weak form, the M\"uller system cancels the hypersingularity already in the strong form through the interior--exterior kernel difference. The structural consequences of
this cancellation for the choice of discretization basis are examined in the following subsection.

\subsection{Hypersingularity Cancellation and Its Consequences for Discretization}

The analytical advantage of the M\"uller formulation is best revealed by expressing the integral operators in tensor form. The most singular contributions in \eqref{eq:mueller_electric} and
\eqref{eq:mueller_magnetic} appear as directional derivatives of the gradient. Applying the tensor identity
\begin{equation}
\big(\bm{j}(\bm{y}) \cdot \nabla\big)\nabla(\varphi_a - \varphi_i)
= \bigl[\nabla\nabla(\varphi_a - \varphi_i)\bigr] \cdot \bm{j}(\bm{y}),
\end{equation}
these terms can be rewritten using the Hessian tensor difference $\nabla\nabla(\varphi_a - \varphi_i)$.

Viewing the M\"uller system merely as a coefficient specialization within the PMCHWT family obscures an essential property of the original construction. The decisive point is not the particular coupling coefficients, but the fact that the exterior and interior Green's kernels
enter exclusively in this difference form. Because their static limits are identical, the leading $\mathcal{O}(R^{-3})$ hypersingularity inherent in the double gradient cancels identically at the operator level, completely circumventing the need to transfer derivatives onto
trial or test functions via weak-form integration by parts.

This distinction has immediate consequences for discretization. In
standard treatments, the apparent presence of second derivatives
motivates regularization through integration by parts, which shifts
the derivatives from the kernel onto the surface-current basis
functions. This strategy is natural in divergence-conforming
discretizations but ties the implementation to basis spaces with
built-in surface-divergence structure. Furthermore, when applied to
the weak form of the M\"uller BIE, it fails to cancel boundary terms,
generating auxiliary line integrals~\cite{Oijala2005}. Treating the
singularity-reduced Hessian kernel difference directly avoids these
complications and enables an isoparametric, high-order nodal Galerkin
discretization without edge-based degrees of freedom or line-integral
corrections.

To preserve this cancellation numerically, the kernel differences are
evaluated using explicit Taylor expansions as $R \to 0$, avoiding the
catastrophic floating-point errors that would arise from subtracting
two nearly equal hypersingular terms.

\subsection{Analytical Cancellation of the Hypersingularity}

Having expressed the singular integral operators in terms of the Hessian
tensor difference, we now address its robust numerical evaluation. Let
$\bm R = \bm{x}-\bm{y}$ denote the distance vector, $R=\|\bm R\|$ its
magnitude, and $\hat{\bm R}=\bm R/R$ the corresponding unit vector. The
scalar Helmholtz kernel admits the Taylor expansion
\begin{equation}
\varphi(R) = \frac{e^{ikR}}{4\pi R} = \frac{1}{4\pi} \sum_{n=0}^{\infty} \frac{(ik)^n}{n!} R^{n-1}.
\end{equation}
Using the differential identity for a radial power function,
\begin{equation}
\nabla\nabla(R^p) = p R^{p-2} \mathbf{I} + p(p-2)R^{p-2} \bigl(\hat{\bm R}\otimes\hat{\bm R}\bigr),
\end{equation}
where $\mathbf I$ is the identity tensor and $\otimes$ the tensor
product, and substituting $p = n-1$, we obtain the Hessian of a single
kernel as
\begin{equation}
\nabla\nabla\varphi(R) = \frac{1}{4\pi} \sum_{n=0}^{\infty} \frac{(ik)^n}{n!} R^{n-3}(n-1) \Bigl[ \mathbf{I} + (n-3)\bigl(\hat{\bm R}\otimes\hat{\bm R}\bigr) \Bigr].
\label{eq:hessian_infinite_series}
\end{equation}
Defining the Hessian tensor difference
\begin{equation}
\mathbf{H}_{\mathrm{diff}}(R) = \nabla\nabla\varphi_a(R) - \nabla\nabla\varphi_i(R),
\end{equation}
and substituting \eqref{eq:hessian_infinite_series} yields
\begin{equation}
\mathbf{H}_{\mathrm{diff}}(R) = \frac{1}{4\pi} \sum_{n=2}^{\infty}\frac{i^n(k_a^n - k_i^n)}{n!} R^{n-3}(n-1)
\Bigl[ \mathbf{I} + (n-3)\bigl(\hat{\bm R}\otimes\hat{\bm R}\bigr) \Bigr].
\label{eq:Hdiff_infinite}
\end{equation}
Equation~\eqref{eq:Hdiff_infinite} furnishes an analytical proof of the
singularity reduction underlying the M\"uller formulation. The $n=0$
term vanishes because $k_a^0 - k_i^0 = 0$, eliminating the
$\mathcal{O}(R^{-3})$ hypersingularity. The $n=1$ term vanishes through the prefactor $n-1 = 0$. The leading singular
behavior of the tensor difference is therefore reduced to the weak
$\mathcal{O}(R^{-1})$ singularity at $n=2$.

In practice, retaining the first five non-vanishing terms ($n=2$ through
$n=6$) of \eqref{eq:Hdiff_infinite} provides sufficient accuracy for
evaluating the singular integrals:
\begin{equation}
\begin{aligned}
\mathbf{H}_{\mathrm{diff}}(R)
\approx\;&
\frac{k_a^2-k_i^2}{8\pi R} \left( \hat{\bm R}\otimes\hat{\bm R}-\mathbf I \right)
-\frac{i(k_a^3-k_i^3)}{12\pi}\,\mathbf I
\\[4pt]
&\;
+\frac{k_a^4-k_i^4}{32\pi}\,R \left( \hat{\bm R}\otimes\hat{\bm R}+\mathbf I \right)\\
&+\frac{i(k_a^5-k_i^5)}{120\pi}\,R^2 \left( 2\hat{\bm R}\otimes\hat{\bm R}+\mathbf I \right)
\\[4pt]
&\;
-\frac{k_a^6-k_i^6}{576\pi}\,R^3 \left( 3\hat{\bm R}\otimes\hat{\bm R}+\mathbf I \right).
\end{aligned}
\label{eq:H_diff_truncated}
\end{equation}
This truncated expansion is employed only when evaluating singular
integrals, where direct floating-point subtraction of two nearly equal
hypersingular kernels would suffer from catastrophic cancellation.


\section{Nodal Orthonormal Frame on High-Order Curved Elements}
\label{sec:manifold_basis}

The nodal Galerkin discretization of the M\"{u}ller system requires a
uniquely defined orthonormal tangent frame
$(\bm{t}_1, \bm{t}_2, \bm{n})$ at each mesh node. This frame
construction is purely local and independent of the global topology of
the scatterer. However, deriving a consistent frame from a high-order isoparametric
mesh poses numerical challenges: standard area-weighted averaging of
adjacent element normals introduces geometric noise that scales with
local element skewness. To achieve geometric fidelity without such artifacts, the construction
below generalizes Max's optimal vertex-angle weights~\cite{Max1999},
originally derived for planar facets approximating a sphere, to
high-order curved elements via a covariant metric formulation.

\subsection{Metric-Weighted Normal Estimation}

Let $v$ be a global node shared by the set of adjacent elements
$\mathcal{E}_v$, and consider a single element $e \in \mathcal{E}_v$.
Let $(\xi_v, \eta_v)$ denote the local parametric coordinates of $v$
within $e$, and let
$\bm{a}_1 = \partial_\xi \bm{x}(\xi_v, \eta_v)$ and
$\bm{a}_2 = \partial_\eta \bm{x}(\xi_v, \eta_v)$ be the covariant
tangent vectors at $v$. All geometric quantities below are evaluated
at element $e$ unless otherwise stated. The metric components
$g_{kk} = \bm{a}_k \cdot \bm{a}_k$ and the surface Jacobian
$J = \|\bm{a}_1 \times \bm{a}_2\|$ follow directly.

Max's original formulation applies to planar elements and uses the
edge lengths and the facet angle at the vertex. To extend this
principle to curved elements, the edge lengths are replaced by the
norms of the covariant tangent vectors, giving the generalized weight
\begin{equation}
w = \frac{\sin\theta}{\|\bm{a}_1\|\,\|\bm{a}_2\|}
  = \frac{J}{g_{11}\,g_{22}},
\end{equation}
where $\theta$ is the angle subtended at $v$ within the element. The
weighted normal contribution is then
\begin{equation}
\bm{U} = w\,\hat{\bm{n}}
= \frac{J}{g_{11}\,g_{22}}
  \cdot\frac{\bm{a}_1 \times \bm{a}_2}{J}
= \frac{\bm{a}_1 \times \bm{a}_2}{g_{11}\,g_{22}}.
\label{eq:U_contribution}
\end{equation}

The Jacobian $J$ cancels exactly in~\eqref{eq:U_contribution}. This
reflects Max's observation that the original discrete formula can be
evaluated without square roots or trigonometric functions, and the
property extends directly to curved manifolds. The final evaluation
requires only one cross product and two dot products.

This metric denominator also acts as an implicit regularizer against
severe parametric distortions. By the Gram identity
$g_{11}\,g_{22} = J^2 + g_{12}^2$, where
$g_{12} = \bm{a}_1 \cdot \bm{a}_2$ is the covariant shear metric, the
magnitude of the weighted contribution takes the form
\begin{equation}
\|\bm{U}\| = \frac{J}{J^2 + g_{12}^2}.
\label{eq:tikhonov_regularization}
\end{equation}
This structure simultaneously satisfies two competing requirements. For
well-proportioned elements ($g_{12} \approx 0$), it reduces to an
inverse-area weight $1/J$, appropriately prioritizing fine-scale local
proximity. For severely skewed sliver elements ($J \to 0$ but
$g_{12}^2 \gg 0$), the shear penalty $g_{12}^2$ automatically drives
the contribution to zero, suppressing geometric noise from high
aspect-ratio elements without ad~hoc thresholding. A
differential-geometric proof of the stability and $\mathcal{O}(h^2)$
convergence of this scheme is given in
Appendix~\ref{app:covariant_normal}.

The nodal normal is assembled by summing over all adjacent elements:
\begin{equation}
\bm{n}_v
  = \frac{\displaystyle\sum_{e\in\mathcal{E}_v}\bm{U}_e}
         {\displaystyle\left\|
           \sum_{e\in\mathcal{E}_v}\bm{U}_e
          \right\|},
\end{equation}
where $\bm{U}_e$ denotes the contribution~\eqref{eq:U_contribution}
evaluated at element $e$.

\subsection{Tangent Frame Completion}

Given $\bm{n}_v$, the two tangent vectors are constructed via a cross
product with a reference axis. A single fixed reference fails when
$\bm{n}_v$ is nearly parallel to that axis, so an adaptive reference
vector is defined as
\begin{equation}
\bm{v}_{\mathrm{ref}} =
\begin{cases}
\hat{\bm{x}}, & \text{if } |\bm{n}_v\cdot\hat{\bm{z}}| > 0.9, \\
\hat{\bm{z}}, & \text{otherwise.}
\end{cases}
\end{equation}
The orthonormal tangent vectors are then
\begin{align}
\bm{t}_{1,v} &= \frac{\bm{v}_{\mathrm{ref}}\times\bm{n}_v}
                      {\|\bm{v}_{\mathrm{ref}}\times\bm{n}_v\|},
\label{eq:t1} \\
\bm{t}_{2,v} &= \bm{n}_v\times\bm{t}_{1,v}.
\label{eq:t2}
\end{align}
The triad $(\bm{t}_{1,v},\bm{t}_{2,v},\bm{n}_v)$ is orthonormal by
construction and is computed once per mesh node.

\subsection{Continuous Frame Interpolation}
\label{sec:frame_interpolation}

The nodal frame vectors $\bm{t}_{\beta,k}$ constructed above are
defined only at the mesh nodes. During assembly, a continuous
orthonormal frame $\hat{\bm{t}}_\beta(\bm{y})$ is required at every
quadrature point $\bm{y}$. The nodal vectors are first interpolated
isoparametrically:
\begin{equation}
\tilde{\bm{t}}_\beta(\bm{y})
  = \sum_{k=1}^6 N_k(\bm{y})\,\bm{t}_{\beta,k},
  \qquad \beta \in \{1,2\},
\end{equation}
where $N_k$ are the $\mathrm{P}_2$ shape functions. Because the interpolated vectors may deviate slightly from the exact
tangent plane, they are orthogonalized against the exact surface
normal at $\bm{y}$, obtained from the isoparametric mapping as
\begin{equation}
\bm{n}(\bm{y})
  = \frac{\bm{a}_1(\bm{y}) \times \bm{a}_2(\bm{y})}
         {\|\bm{a}_1(\bm{y}) \times \bm{a}_2(\bm{y})\|},
\end{equation}
where $\bm{a}_1(\bm{y}) = \partial_\xi \bm{x}$ and
$\bm{a}_2(\bm{y}) = \partial_\eta \bm{x}$ are the covariant tangent
vectors at $\bm{y}$. Introducing the tangent-plane projector
$\mathbf{P}(\bm{y}) = \mathbf{I}
- \bm{n}(\bm{y})\otimes\bm{n}(\bm{y})$, the Gram--Schmidt procedure
yields the orthonormal frame vectors
\begin{align}
\hat{\bm{t}}_1(\bm{y})
  &= \frac{\mathbf{P}\,\tilde{\bm{t}}_1}
          {\|\mathbf{P}\,\tilde{\bm{t}}_1\|},
\\[4pt]
\hat{\bm{t}}_2(\bm{y})
  &= \frac{\mathbf{P}\,\tilde{\bm{t}}_2
           - \bigl(\hat{\bm{t}}_1 \cdot
             \mathbf{P}\,\tilde{\bm{t}}_2\bigr)\hat{\bm{t}}_1}
          {\|\mathbf{P}\,\tilde{\bm{t}}_2
           - \bigl(\hat{\bm{t}}_1 \cdot
             \mathbf{P}\,\tilde{\bm{t}}_2\bigr)\hat{\bm{t}}_1\|},
\end{align}
where the argument $(\bm{y})$ has been suppressed for brevity. This construction ensures exact tangency $\hat{\bm{t}}_\beta(\bm{y}) \cdot \bm{n}(\bm{y}) \equiv 0$ everywhere on the curved manifold.

\section{Nodal Galerkin Discretization and Singularity Extraction}
\label{sec:scalarization}

In this section, we apply Galerkin testing to the M\"uller
system~\eqref{eq:mueller_electric}--\eqref{eq:mueller_magnetic} using
vector basis and test functions formed as products of scalar nodal
shape functions and the continuous orthonormal tangent frame derived
in Section~\ref{sec:frame_interpolation}. This reduces the tensorial
integral operators to scalar kernel blocks, from which the singular
principal parts are then isolated for the Sauter--Schwab quadrature.

\subsection{Operator $\mathcal{K}_1$}

In \eqref{eq:mueller_electric} and \eqref{eq:mueller_magnetic}, the
integral operator $\mathcal{K}_1$ acting on a generic tangential vector
field $\bm{v}(\bm{y})$ is
\begin{equation}
\mathcal{K}_1[\bm{v}](\bm{x})= \int_\Gamma \Big\{
  \bigl(k_a^2\varphi_a - k_i^2\varphi_i\bigr)
  \bigl[\bm{n}(\bm{x})\times\bm{v}(\bm{y})\bigr]
  + \bm{n}(\bm{x})\times  \bigl[\mathbf{H}_{\mathrm{diff}}\cdot\bm{v}(\bm{y})\bigr]
\Big\}\,d\Gamma_y.
\label{eq:K1_operator}
\end{equation}

To discretize the M\"uller BIEs, the continuous surface currents are
projected onto a finite-dimensional tangential trial space
$\mathcal{V}_h$. The electric surface current $\bm{j}(\bm{y})$ is
approximated by
\begin{equation}
\bm{j}_h(\bm{y}) = \sum_{n=1}^{N_{\mathrm{dof}}} c_n\,
\bm{\Lambda}_n(\bm{y}),
\label{eq:current_expansion}
\end{equation}
where $\{\bm{\Lambda}_n\}_{n=1}^{N_{\mathrm{dof}}}$ is the basis for
$\mathcal{V}_h$ and $c_n$ are the unknown expansion coefficients. Each tangential vector basis function is defined as
\begin{equation}
\bm{\Lambda}_{n(j,\beta)}(\bm{y})
  = N_j(\bm{y})\,\hat{\bm{t}}_\beta(\bm{y}),
\end{equation}
coupling the scalar nodal shape function $N_j(\bm{y})$ with the
continuous orthonormal frame vector $\hat{\bm{t}}_\beta(\bm{y})$.
The $\mathrm{P}_2$ shape functions $N_j$ serve purely as scalar
interpolants for the current amplitudes, while
$\hat{\bm{t}}_\beta$ governs the vector orientation. The magnetic
surface current $\bm{j}_m(\bm{y})$ is expanded analogously.

Testing with
$\bm{W}_i^\alpha(\bm{x}) = N_i(\bm{x})\,\hat{\bm{t}}_\alpha(\bm{x})$
and applying the identity
$\bm{A}\cdot(\bm{B}\times\bm{C}) = (\bm{A}\times\bm{B})\cdot\bm{C}$
with the rotated test vector
$\hat{\bm{s}}_\alpha(\bm{x})
= \hat{\bm{t}}_\alpha(\bm{x})\times\bm{n}(\bm{x})$
reduces the integrand to the scalar kernel
\begin{equation}
\kappa_{1,\alpha\beta}(\bm{x},\bm{y})
  = \bigl(k_a^2\varphi_a - k_i^2\varphi_i\bigr)\,
    \bigl(\hat{\bm{s}}_\alpha(\bm{x})\cdot
          \hat{\bm{t}}_\beta(\bm{y})\bigr)
  + \hat{\bm{s}}_\alpha(\bm{x})\cdot
    \bigl(\mathbf{H}_{\mathrm{diff}}\cdot
          \hat{\bm{t}}_\beta(\bm{y})\bigr).
\label{eq:kappa1_alphabeta}
\end{equation}
All cross products have been eliminated; the integrand is expressed entirely through scalar dot products of the local orthonormal
frames. The corresponding Galerkin matrix entry is
\begin{equation}
\bigl[\mathbf{K}_{1,\alpha\beta}\bigr]_{ij}
  = \int_\Gamma N_i(\bm{x}) \int_\Gamma
    \kappa_{1,\alpha\beta}(\bm{x},\bm{y})\,
    N_j(\bm{y})\,d\Gamma_y\,d\Gamma_x.
\label{eq:K1_matrix_entry}
\end{equation}

For regular integrations over disjoint elements, the exact Hessian
difference $\mathbf{H}_{\mathrm{diff}}$ is evaluated as
\begin{equation}
\mathbf{H}_{\mathrm{diff}}(R)
  = \bigl[f_{k_a}(R) - f_{k_i}(R)\bigr]\,
    \hat{\bm{R}}\otimes\hat{\bm{R}}
  - \bigl[g_{k_a}(R) - g_{k_i}(R)\bigr]\,\mathbf{I},
\label{eq:Hdiff_exact}
\end{equation}
where $f_k$ and $g_k$ are
\begin{equation}
\begin{aligned}
f_k(R) &= \frac{e^{ikR}}{4\pi R^3}\bigl(3 - 3ikR - k^2 R^2\bigr), \\[4pt]
g_k(R) &= \frac{e^{ikR}}{4\pi R^3}\bigl(1 - ikR\bigr).
\end{aligned}
\end{equation}

For singular integrations, direct evaluation of~\eqref{eq:Hdiff_exact}
is avoided to prevent catastrophic cancellation. Instead, the leading
term of the expansion~\eqref{eq:Hdiff_infinite} is isolated and
integrated by Sauter--Schwab quadrature. Substituting this term
into~\eqref{eq:kappa1_alphabeta} yields the singular kernel
\begin{equation}
\kappa_{1,\alpha\beta}^{\mathrm{s}}(\bm{x},\bm{y})
  = \frac{k_a^2 - k_i^2}{8\pi R} \Bigl[
      \hat{\bm{s}}_\alpha(\bm{x})\cdot\hat{\bm{t}}_\beta(\bm{y})
  + \bigl(\hat{\bm{s}}_\alpha(\bm{x})\cdot\hat{\bm{R}}\bigr)
        \bigl(\hat{\bm{R}}\cdot\hat{\bm{t}}_\beta(\bm{y})\bigr)
    \Bigr].
\label{eq:kappa_S}
\end{equation}
The regular remainder
$\kappa_{1,\alpha\beta} - \kappa_{1,\alpha\beta}^{\mathrm{s}}$
is evaluated by the 12-point Witherden--Vincent rule \cite{Witherden2015} used for the far-field integrations.

\subsection{Operator $\mathcal{K}_2^{\chi}$}

To treat the cross-product integral operators in a unified manner, we
introduce the material label $\chi \in \{\varepsilon, \mu\}$ and the
composite Helmholtz Green's function
$\Phi_\chi = \chi_a\varphi_a - \chi_i\varphi_i$. The operator
$\mathcal{K}_2^\chi$ acting on a generic tangential field
$\bm{v}(\bm{y})$ reads
\begin{equation}
\mathcal{K}_2^\chi[\bm{v}](\bm{x})
  = \int_\Gamma \bm{n}(\bm{x})\times
    \bigl(\bm{v}(\bm{y})\times\nabla_{\bm{x}}\Phi_\chi\bigr)
    \,d\Gamma_y.
\end{equation}
In the M\"uller system~\eqref{eq:mueller_electric}
and~\eqref{eq:mueller_magnetic}, $\mathcal{K}_2^\chi$ is applied to
the magnetic current $\bm{j}_m$ with $\chi = \varepsilon$, and to the
electric current $\bm{j}$ with $\chi = \mu$.

Expanding $\bm{v}(\bm{y})$ in the orthonormal nodal basis and testing
with
$\bm{W}_i^\alpha(\bm{x}) = N_i(\bm{x})\,\hat{\bm{t}}_\alpha(\bm{x})$,
the identity
$\bm{A}\times(\bm{B}\times\bm{C})
= \bm{B}(\bm{A}\cdot\bm{C}) - \bm{C}(\bm{A}\cdot\bm{B})$ yields the scalar kernel
\begin{equation}
\kappa_{2,\alpha\beta}^\chi(\bm{x},\bm{y})
  = \bigl[\hat{\bm{t}}_\alpha(\bm{x})\cdot
          \hat{\bm{t}}_\beta(\bm{y})\bigr]
    \bigl(\bm{n}(\bm{x})\cdot\nabla_x\Phi_\chi\bigr)
   - \bigl[\hat{\bm{t}}_\alpha(\bm{x})\cdot\nabla_x\Phi_\chi\bigr]
    \bigl(\bm{n}(\bm{x})\cdot\hat{\bm{t}}_\beta(\bm{y})\bigr).
\label{eq:kappa2_alpha_beta}
\end{equation}
The corresponding Galerkin matrix entry is
\begin{equation}
\bigl[\mathbf{K}_{2,\alpha\beta}^{\chi}\bigr]_{ij}
  = \int_\Gamma N_i(\bm{x}) \int_\Gamma
    \kappa_{2,\alpha\beta}^\chi(\bm{x},\bm{y})\,
    N_j(\bm{y})\,d\Gamma_y\,d\Gamma_x.
\label{eq:K2_matrix_entry}
\end{equation}

As in the preceding subsection, the singular part must be isolated for the Sauter--Schwab quadrature. Expanding the gradient of the Helmholtz kernel gives
\begin{equation}
\nabla_{\bm{x}}\varphi_{a,i}(R)
  = \frac{1}{4\pi}\!\left(-\frac{1}{R^2} - \frac{k_{a,i}^2}{2}
    + \mathcal{O}(R)\right)\hat{\bm{R}}.
\end{equation}
Hence, the leading singularity of $\nabla_{\bm{x}}\Phi_\chi$ is
$\mathcal{O}(R^{-2})$, with the singular part
\begin{equation}
\bigl(\nabla_{\bm{x}}\Phi_\chi\bigr)^{\mathrm{s}}
  = -\frac{\chi_a - \chi_i}{4\pi R^2}\,\hat{\bm{R}}.
\end{equation}
Substituting into~\eqref{eq:kappa2_alpha_beta} yields the singular
kernel
\begin{equation}
\kappa_{2,\alpha\beta}^{\chi,\mathrm{s}}
  = -\frac{\chi_a-\chi_i}{4\pi R^2} \Bigl\{
      \bigl[\hat{\bm{t}}_\alpha(\bm{x})\cdot
            \hat{\bm{t}}_\beta(\bm{y})\bigr]
      \bigl(\bm{n}(\bm{x})\cdot\hat{\bm{R}}\bigr)
   -\bigl[\hat{\bm{t}}_\alpha(\bm{x})\cdot\hat{\bm{R}}\bigr]
      \bigl(\bm{n}(\bm{x})\cdot\hat{\bm{t}}_\beta(\bm{y})\bigr) \Bigr\}.
\label{eq:kappa2_S}
\end{equation}
Although~\eqref{eq:kappa2_S} exhibits a nominal $\mathcal{O}(R^{-2})$
prefactor, the geometric structure of the bracket reduces the
effective singularity to $\mathcal{O}(R^{-1})$ on smooth surfaces.
This reduction follows from the asymptotic behavior of the two
projections in the bracket, which we analyze in turn.

\emph{First projection, $\bm{n}(\bm{x})\cdot\hat{\bm{R}}$.}
Introduce Riemann normal coordinates (RNC) $u^\mu$
($\mu \in \{1,2\}$) centered at $\bm{x}$~\cite{doCarmo1976}, with
Einstein summation implied over repeated indices. Locally parameterize
the surface as $\bm{r}(u^1,u^2)$ with covariant basis
$\bm{a}_\mu = \partial_\mu\bm{r}$. The Gauss formula reads
$\partial_\mu\partial_\nu\bm{r}
= \Gamma^\sigma_{\mu\nu}\bm{a}_\sigma + L_{\mu\nu}\bm{n}$,
where $\Gamma^\sigma_{\mu\nu}$ are the Christoffel symbols and
$L_{\mu\nu}$ the second fundamental form. At the RNC origin the
Christoffel symbols vanish, so expanding the distance vector
$\bm{R} = \bm{x} - \bm{y}$ gives
\begin{equation}
\bm{n}(\bm{x})\cdot\hat{\bm{R}}
  = -\frac{1}{2R}\,L_{\mu\nu}(\bm{x})\,u^\mu u^\nu
    + \mathcal{O}(R^2)
  = \mathcal{O}(R).
\end{equation}

\emph{Second projection,
$\bm{n}(\bm{x})\cdot\hat{\bm{t}}_\beta(\bm{y})$.}
The Gram--Schmidt orthogonalization of
Section~\ref{sec:frame_interpolation} ensures exact tangency,
$\bm{n}(\bm{y})\cdot\hat{\bm{t}}_\beta(\bm{y}) \equiv 0$. By the
Weingarten equation $\partial_\mu\bm{n} = -L^\nu_\mu\bm{a}_\nu$, the
normal field expands smoothly as
\begin{equation}
\bm{n}(\bm{x})
  = \bm{n}(\bm{y}) + \partial_\mu\bm{n}(\bm{y})\,u^\mu
    + \mathcal{O}(R^2)
  = \bm{n}(\bm{y}) + \mathcal{O}(R),
\end{equation}
so that
\begin{equation}
\bm{n}(\bm{x})\cdot\hat{\bm{t}}_\beta(\bm{y})
  = \bm{n}(\bm{y})\cdot\hat{\bm{t}}_\beta(\bm{y})
    + \mathcal{O}(R)
  = \mathcal{O}(R).
\end{equation}

Both projections therefore scale as $\mathcal{O}(R)$, reducing each
term in~\eqref{eq:kappa2_S} to $\mathcal{O}(R^{-1})$. Together with
the earlier analysis of $\mathcal{K}_1$, this confirms that both
$\kappa_{1,\alpha\beta}^{\mathrm{s}}$ and
$\kappa_{2,\alpha\beta}^{\chi,\mathrm{s}}$ are at most weakly
singular.

These singular kernels are integrated by Sauter--Schwab quadrature, which handles the three topological singular classes (common face, common edge, and common vertex) in a
unified manner. Duffy-type transformations map each singular surface
double integral to a regular integral over the unit hypercube
$[0,1]^4$, producing a tensor-product integrand amenable to standard
Gauss--Legendre quadrature. No analytical extraction of the singular
core is required, so the scheme applies directly to high-order
isoparametric elements.

\subsection{Assembled Linear System}

Combining the scalar kernel blocks derived in the preceding subsections,
the Galerkin discretization of the M\"uller
system~\eqref{eq:mueller_electric}--\eqref{eq:mueller_magnetic} yields
the global linear system
\begin{equation}
\begin{bmatrix}
\dfrac{i}{\omega}\,\mathbf{K}_1
  & \dfrac{\varepsilon_i+\varepsilon_a}{2}\,\mathbf{M}
    + \mathbf{K}_2^\varepsilon \\[8pt]
\dfrac{\mu_i+\mu_a}{2}\,\mathbf{M}
    + \mathbf{K}_2^\mu
  & -\dfrac{i}{\omega}\,\mathbf{K}_1
\end{bmatrix}
\begin{bmatrix}
\bm{c}_j \\[4pt] \bm{c}_{j_m}
\end{bmatrix}
=
\begin{bmatrix}
\bm{b}_E \\[4pt] \bm{b}_H
\end{bmatrix},
\label{eq:global_system}
\end{equation}
where $\bm{c}_j$ and $\bm{c}_{j_m}$ are the vectors of unknown
expansion coefficients for the electric and magnetic surface currents,
respectively.

The operator matrices $\mathbf{K}_1$ and $\mathbf{K}_2^\chi$ are
assembled from the scalar kernel entries~\eqref{eq:K1_matrix_entry}
and~\eqref{eq:K2_matrix_entry}. The Galerkin mass matrix
$\mathbf{M}$ collects the identity contributions
$\tfrac{\varepsilon_i + \varepsilon_a}{2}\,\bm{j}_m$ and
$\tfrac{\mu_i + \mu_a}{2}\,\bm{j}$ from the left-hand sides
of~\eqref{eq:mueller_electric} and~\eqref{eq:mueller_magnetic}.
Because the test and trial basis functions share the same orthonormal
tangent frame, $\mathbf{M}$ is block-diagonal in the component indices
$(\alpha,\beta)$, with each diagonal block given by the scalar
$\mathrm{P}_2$ Gram matrix:
\begin{equation}
[\mathbf{M}_{\alpha\beta}]_{ij}
  = \delta_{\alpha\beta}
    \int_\Gamma N_i(\bm{x})\,N_j(\bm{x})\,d\Gamma.
\label{eq:mass_matrix}
\end{equation}

The RHS vectors are obtained by testing the incident fields
against the rotated frame vectors
$\hat{\bm{s}}_\alpha(\bm{x})
= \hat{\bm{t}}_\alpha(\bm{x})\times\bm{n}(\bm{x})$:
\begin{align}
[\bm{b}_E]_{(\alpha,i)}
  &= - \varepsilon_a \int_\Gamma N_i(\bm{x})\,
     \hat{\bm{s}}_\alpha(\bm{x})\cdot
     \bm{E}^{\mathrm{inc}}(\bm{x})\,d\Gamma,
\label{eq:rhs_E} \\[4pt]
[\bm{b}_H]_{(\alpha,i)}
  &= \mu_a \int_\Gamma N_i(\bm{x})\,
     \hat{\bm{s}}_\alpha(\bm{x})\cdot
     \bm{H}^{\mathrm{inc}}(\bm{x})\,d\Gamma.
\label{eq:rhs_H}
\end{align}
These arise from the excitation terms
$\bm{j}_m^{\mathrm{inc}}$ and $\bm{j}^{\mathrm{inc}}$
in~\eqref{eq:mueller_electric} and~\eqref{eq:mueller_magnetic} after the same reduction used for the operator kernels.

\section{Morton-Ordered Block Jacobi Preconditioning}
\label{sec:preconditioning}

The Galerkin discretization of Section~\ref{sec:scalarization} yields
a dense linear system
\begin{equation}
\mathbf{A}\,\bm{x} = \bm{b},
\qquad \mathbf{A} \in \mathbb{C}^{N\times N},
\end{equation}
which is solved iteratively by GMRES. The Fredholm second-kind
structure of the M\"uller formulation keeps the condition number of
$\mathbf{A}$ bounded as the mesh parameter $h \to 0$, so that the
GMRES iteration count does not grow under mesh refinement. This
$h$-independence does not, however, translate into rapid convergence
in all scattering regimes. Extreme material parameters, complex
geometries, and elevated electrical sizes can each broaden the
spectrum of the discrete operator, disrupting the diagonal dominance
of $\mathbf{A}$ and inflating the Krylov subspace dimension.

To address this spectral broadening, we introduce a Morton-ordered
Block Jacobi (MBJ) preconditioner. Rather than relying on hierarchical
tree structures or \emph{a~posteriori} algebraic permutations, the MBJ
strategy performs spatial clustering \emph{a~priori} at the mesh
preprocessing stage. Immediately after loading the surface mesh and
before any operator assembly, a double Morton permutation (Z-order
space-filling curve) is applied. The first pass reorders the global
nodes, mapping spatially proximate degrees of freedom to contiguous
memory indices. The second pass reorders the surface elements for
cache locality during quadrature assembly. This preprocessing maps
three-dimensional spatial proximity to one-dimensional index
contiguity, so that $\mathbf{A}$ is assembled with its strongest
near-field couplings concentrated along the main diagonal.

The contiguous node index set is partitioned into $N_b$ disjoint
sequential clusters $\mathcal{I}_m$, each containing $B$ nodes. For
each cluster, the local preconditioner block is assembled as the
$2B \times 2B$ coupled system
\begin{equation}
\mathbf{A}_m =
\begin{bmatrix}
\dfrac{i}{\omega}\,\mathbf{K}_{1,m}
  & \dfrac{\varepsilon_i+\varepsilon_a}{2}\,\mathbf{M}_m
    + \mathbf{K}_{2,m}^\varepsilon \\[8pt]
\dfrac{\mu_i+\mu_a}{2}\,\mathbf{M}_m
    + \mathbf{K}_{2,m}^\mu
  & -\dfrac{i}{\omega}\,\mathbf{K}_{1,m}
\end{bmatrix},
\label{eq:precond_block}
\end{equation}
where $\mathbf{K}_{1,m}$, $\mathbf{K}_{2,m}^\chi$, and $\mathbf{M}_m$
denote the restrictions of the corresponding global operators to the
index set $\mathcal{I}_m$. Each block thus preserves the fully coupled electromagnetic operator structure of~\eqref{eq:global_system}, including
the mass matrix and both off-diagonal interactions, for a spatially
contiguous group of nodes. The MBJ preconditioner is the block-diagonal
matrix
\begin{equation}
\mathbf{P}
  = \operatorname{blockdiag}(\mathbf{A}_1,\mathbf{A}_2,\dots,
    \mathbf{A}_{N_b}).
\end{equation}

Because the Morton reordering guarantees that each diagonal block
captures the dominant local couplings, no indirect memory gathering
or algebraic restriction operators are required. Each local block
$\mathbf{A}_m$ preserves the full $2 \times 2$ operator structure of
the Galerkin system including $\mathbf{M}_m$; since $\mathbf{M}$
acts locally, it is captured within the diagonal blocks and cancels
from the off-diagonal remainder. The remaining off-diagonal
interactions are weakly singular $\mathcal{O}(R^{-1})$ kernels on
spatially separated subdomains, which map $L^2(\Gamma)$
compactly~\cite{Kress2014}. The right-preconditioned operator is
therefore a compact perturbation of the identity, with spectrum
clustered at unity, and GMRES converges
superlinearly~\cite{Nevanlinna1993}.


\section{Numerical Validation}
\label{sec:validation}

\subsection{Implementation and Computational Environment}
\label{sec:implementation}

The solver is implemented in Julia and executed on a laptop equipped
with an AMD Ryzen AI 9 HX 370 processor (8 cores, 16 threads) and
32\,GB LPDDR5 memory. All timings reported below are measured on this
platform.

The three operator matrices $\mathbf{K}_1$,
$\mathbf{K}_2^\varepsilon$, and $\mathbf{K}_2^\mu$ share identical
geometric primitives: inter-element distance, the complex
exponential pair $e^{ik_a R}$ and $e^{ik_i R}$, and the orthonormal
frame dot products. They are therefore assembled in a single fused
pass over the element-pair table. Each quadrature-point evaluation
computes the shared geometry once and distributes the result across
the four tensorial components ($\alpha, \beta \in \{1,2\}$) of the
three operators, yielding twelve scalar accumulations per evaluation.
The frequency-dependent Taylor coefficients of
$\mathbf{H}_{\mathrm{diff}}$ are precomputed outside the innermost
quadrature loops.

Regular integrals over the quadratic elements are evaluated using the
12-point Witherden--Vincent rule. Because the hypersingularity
cancellation of Section~\ref{sec:mueller_formulation} reduces all
singular kernels to at most $\mathcal{O}(R^{-1})$, a Sauter--Schwab
quadrature order of $N = 4$ points per dimension ($4^4 = 256$
evaluations per singular element pair) suffices on curved
$\mathrm{P}_2$ surfaces. The overall dense assembly scales as
$\mathcal{O}(N_{\mathrm{dof}}^2)$.

\subsection{Benchmark~I: Oblique Scattering by a Gold Spheroid}
\label{sec:val_spheroid}

\subsubsection{Problem Setup}

The first benchmark considers a gold prolate spheroid with semi-axes
$a = b = 200$\,nm and $c = 300$\,nm, illuminated at vacuum wavelength
$\lambda = 418.9$\,nm. At this wavelength, the gold permittivity is
$\varepsilon_r = -1.0922 + 5.5139i$ (refractive index
$1.5048 + 1.8321i$)~\cite{Sun2017}, yielding a size parameter $k_a a = 2.9998$. The incident plane wave is parameterized
by Euler angles
$(\theta_p, \phi_p, \alpha_p) = (45^\circ, 30^\circ, 60^\circ)$,
which breaks all geometric and polarization symmetries. Semi-analytical
reference values are provided by
SMARTIES~\cite{Somerville2016}, a freely available solver that computes
the electromagnetic response of spheroids via the Extended Boundary
Condition Method (EBCM)~\cite{Waterman1971, Mishchenko1996,
Mishchenko2002}. The EBCM expands the fields in vector spherical wave
functions and enforces the extinction theorem, converging spectrally
with respect to the expansion order. For the present geometry, the
EBCM solution is accurate to double-precision level at moderate
truncation order.

\subsubsection{Convergence and Far-Field Validation}
\label{sec:spheroid_convergence}

\begin{figure}[htbp]
    \centering
    \includegraphics[width=0.7\columnwidth]{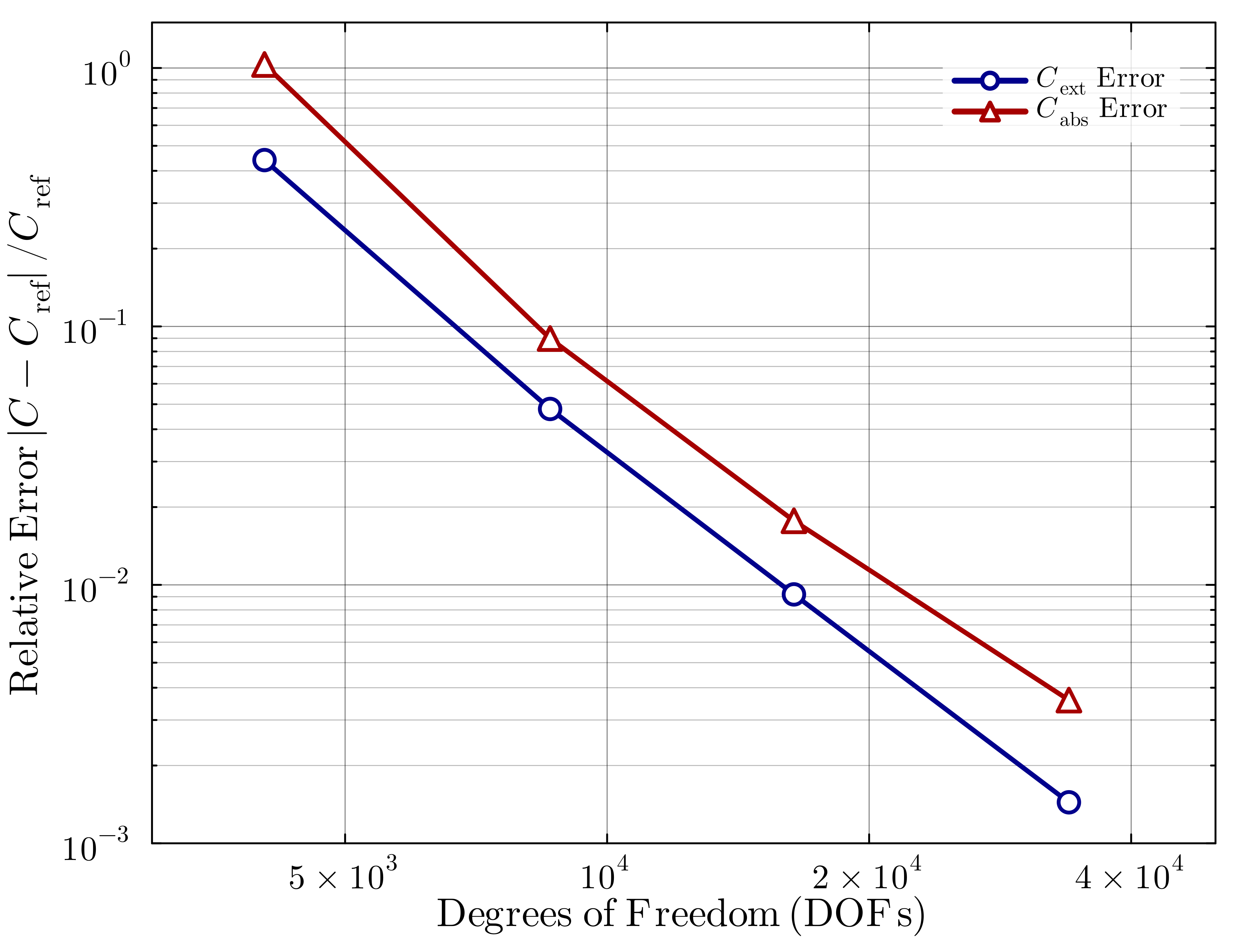}
    \caption{$h$-convergence of the extinction and absorption cross-section errors
for the gold prolate spheroid. Both quantities decrease monotonically
on the log--log scale.}
    \label{fig:h_convergence}
\end{figure}

Four isoparametric $\mathrm{P}_2$ meshes spanning 4\,040 to 33\,928
DOFs (Table~\ref{tab:mesh_topology_spheroid}) are solved under the
identical oblique excitation. Fig.~\ref{fig:h_convergence} plots the
relative errors of the extinction and absorption cross sections on a
log--log scale against the SMARTIES reference. Both quantities converge
monotonically. Measuring the empirical slopes across consecutive mesh
transitions with respect to the characteristic mesh size
($h \sim N_{\mathrm{dof}}^{-1/2}$) yields rates between
$\mathcal{O}(h^{5.1})$ and $\mathcal{O}(h^{5.9})$ for extinction and
between $\mathcal{O}(h^{4.4})$ and $\mathcal{O}(h^{6.5})$ for
absorption. These rates substantially exceed the $\mathcal{O}(h^3)$
convergence expected for the surface currents in a $\mathrm{P}_2$
discretization, consistent with the well-known superconvergence of
far-field cross sections as global integral functionals of the surface
currents. At the finest level, the relative errors for the extinction ($C_{\mathrm{ext}}$), absorption ($C_{\mathrm{abs}}$), and scattering ($C_{\mathrm{sca}}$) cross sections are $0.144\%$, $0.358\%$, and $0.016\%$, respectively.

The energy balance error
$|C_{\mathrm{ext}} - C_{\mathrm{abs}} - C_{\mathrm{sca}}| /
C_{\mathrm{ext}}$ remains below $10^{-12}\%$ (i.e., $\mathcal{O}(10^{-14})$
in relative magnitude) at every mesh level, confirming that the optical
theorem is satisfied to double-precision accuracy independently of the
mesh resolution.

Table~\ref{tab:convergence_performance} summarizes the computational
performance across the refinement sequence. As the system size grows
from 4\,040 to 33\,928 DOFs, the GMRES iteration count remains bounded
within $[106,\,126]$, empirically confirming the mesh-independent
conditioning characteristic of the Fredholm second-kind formulation.

\begin{table}[htbp]
\centering
\caption{Mesh sequence for the spheroid $h$-convergence study.}
\label{tab:mesh_topology_spheroid}
\setlength{\tabcolsep}{8pt}
\begin{tabular}{@{}cccc@{}}
\toprule
\textbf{Level} & \textbf{Nodes} & \textbf{Elements} & \textbf{DOFs} \\
\midrule
M1 & 1\,010 & 504    & 4\,040  \\
M2 & 2\,150 & 1\,074 & 8\,600  \\
M3 & 4\,098 & 2\,048 & 16\,392 \\
M4 & 8\,482 & 4\,240 & 33\,928 \\
\bottomrule
\end{tabular}
\end{table}

\begin{table}[htbp]
\centering
\caption{Computational performance and cross-section errors for the
         gold spheroid $h$-convergence study. Reference: SMARTIES.}
\label{tab:convergence_performance}
\begin{tabular*}{\textwidth}{@{\extracolsep{\fill}}ccccccc@{}}
\toprule
\textbf{Level}
  & \textbf{Assembly (s)} 
  & \textbf{Solve (s)}
  & \textbf{GMRES iters}
  & $\bm{\delta C_{\mathrm{ext}}}$ \textbf{(\%)} 
  & $\bm{\delta C_{\mathrm{abs}}}$ \textbf{(\%)}
  & $\bm{\delta C_{\mathrm{sca}}}$ \textbf{(\%)} \\
\midrule
M1  & 0.78  & 0.67  & 124 & 44.021  & 103.184 & 0.280 \\
M2  & 2.99  & 2.35  & 106 & 4.794   & 8.978   & 1.660 \\
M3  & 10.28 & 9.77  & 114 & 0.919   & 1.770   & 0.281 \\
M4  & 52.79 & 46.75 & 126 & 0.144   & 0.358   & 0.016 \\
\bottomrule
\end{tabular*}
\end{table}

\begin{figure}[htbp]
    \centering
    \includegraphics[width=0.7\columnwidth]{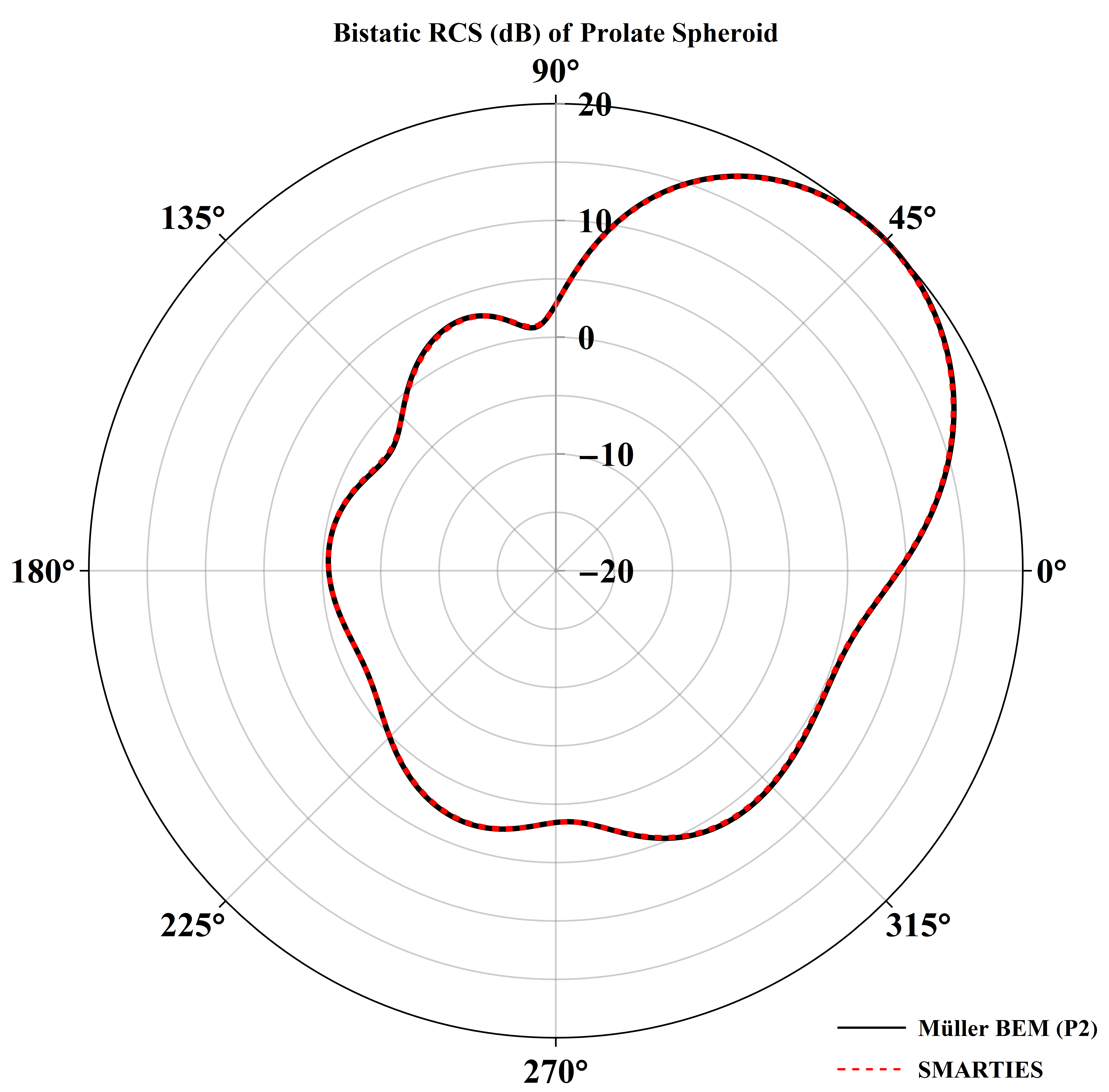}
    \caption{Bistatic RCS on the cut plane $\varphi_s = 30^\circ$ for
             the gold prolate spheroid. Solid: M\"uller BIE
             (5\,162 nodes, 2\,580 $\mathrm{P}_2$ elements). Dashed:
             SMARTIES ($N_{\max}{=}15$).}
    \label{fig:rcs_spheroid}
\end{figure}

Cross sections confirm global energy accuracy but do not verify the
angular distribution of the scattered field. To assess the latter,
Fig.~\ref{fig:rcs_spheroid} compares the bistatic RCS on the cut
plane $\varphi_s = 30^\circ$ for a mesh of 5\,162 nodes
(20\,648 DOFs) against a SMARTIES reference computed with multipole
truncation order $N_{\max} = 15$. The M\"uller BIE pattern matches
the reference over the full angular range, reproducing both the
forward-hemisphere main lobe and the broad minima in the back
hemisphere.

\subsection{Benchmark~II: Silver Spheroid LSPR Spectrum}

\subsubsection{Problem Setup}

The second benchmark evaluates the framework under the strongly
dispersive permittivity characteristic of plasmonic resonance. At optical frequencies, noble metals such as silver exhibit a negative real
permittivity whose magnitude varies rapidly with wavelength. Near the LSPR, the real part of $\varepsilon_r$ reaches large negative values while the imaginary
part remains small, producing strong field enhancement at the particle
surface and placing stringent demands on the accuracy and conditioning
of the numerical solver.

\begin{figure}[htbp]
    \centering
    \includegraphics[width=0.6\columnwidth, height=0.35\textheight,
      keepaspectratio]{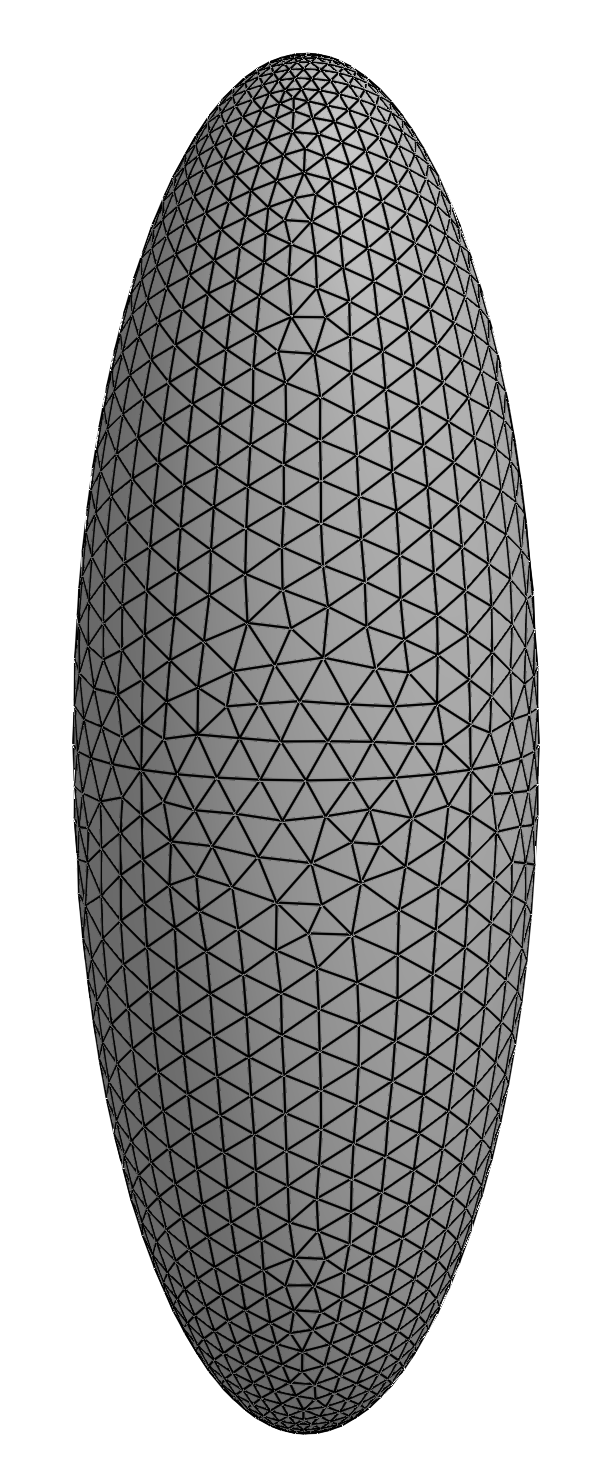}
    \caption{Surface discretization of the $30 \times 90$\,nm silver
             prolate spheroid (3\,072 isoparametric $\mathrm{P}_2$
             elements).}
    \label{fig:lspr_mesh}
\end{figure}

A silver prolate spheroid with semi-axes 15\,nm (transverse) and
45\,nm (longitudinal) is illuminated in free space by a plane wave propagating perpendicular to the major axis with electric-field
polarization along it, exciting the longitudinal dipole resonance. The complex
permittivity of silver is interpolated from the experimental
measurements of Johnson and Christy~\cite{Johnson1972}
(Table~\ref{tab:jc_ag_data}). The mesh comprises 3\,072 $\mathrm{P}_2$
elements and 6\,146 nodes (24\,584 DOFs), shown in
Fig.~\ref{fig:lspr_mesh}.

\begin{table}[htbp]
    \centering
    \caption{Optical constants of silver from Johnson and
             Christy~\cite{Johnson1972}. The relative permittivity is
             $\varepsilon_r = (n + i\kappa)^2$.}
    \label{tab:jc_ag_data}
    \setlength{\tabcolsep}{8pt}
    \begin{tabular}{@{}ccc|ccc@{}}
    \toprule
    $\lambda$ (nm) & $n$ & $\kappa$
      & $\lambda$ (nm) & $n$ & $\kappa$ \\
    \midrule
    300.9 & 1.34 & 0.964 & 450.9 & 0.04 & 2.657 \\
    310.7 & 1.13 & 0.616 & 471.4 & 0.05 & 2.869 \\
    320.4 & 0.81 & 0.392 & 495.9 & 0.05 & 3.093 \\
    331.5 & 0.17 & 0.829 & 520.9 & 0.05 & 3.324 \\
    342.5 & 0.14 & 1.142 & 548.6 & 0.06 & 3.586 \\
    354.2 & 0.10 & 1.419 & 582.1 & 0.05 & 3.858 \\
    367.9 & 0.07 & 1.657 & 616.8 & 0.06 & 4.152 \\
    381.5 & 0.05 & 1.864 & 659.5 & 0.05 & 4.483 \\
    397.4 & 0.05 & 2.070 & 704.5 & 0.04 & 4.838 \\
    413.3 & 0.05 & 2.275 & 756.0 & 0.03 & 5.242 \\
    430.5 & 0.04 & 2.462 & 821.1 & 0.04 & 5.727 \\
    \bottomrule
    \end{tabular}
\end{table}

\subsubsection{LSPR Spectrum}

\begin{figure}[htbp]
    \centering
    \includegraphics[width=0.7\columnwidth]{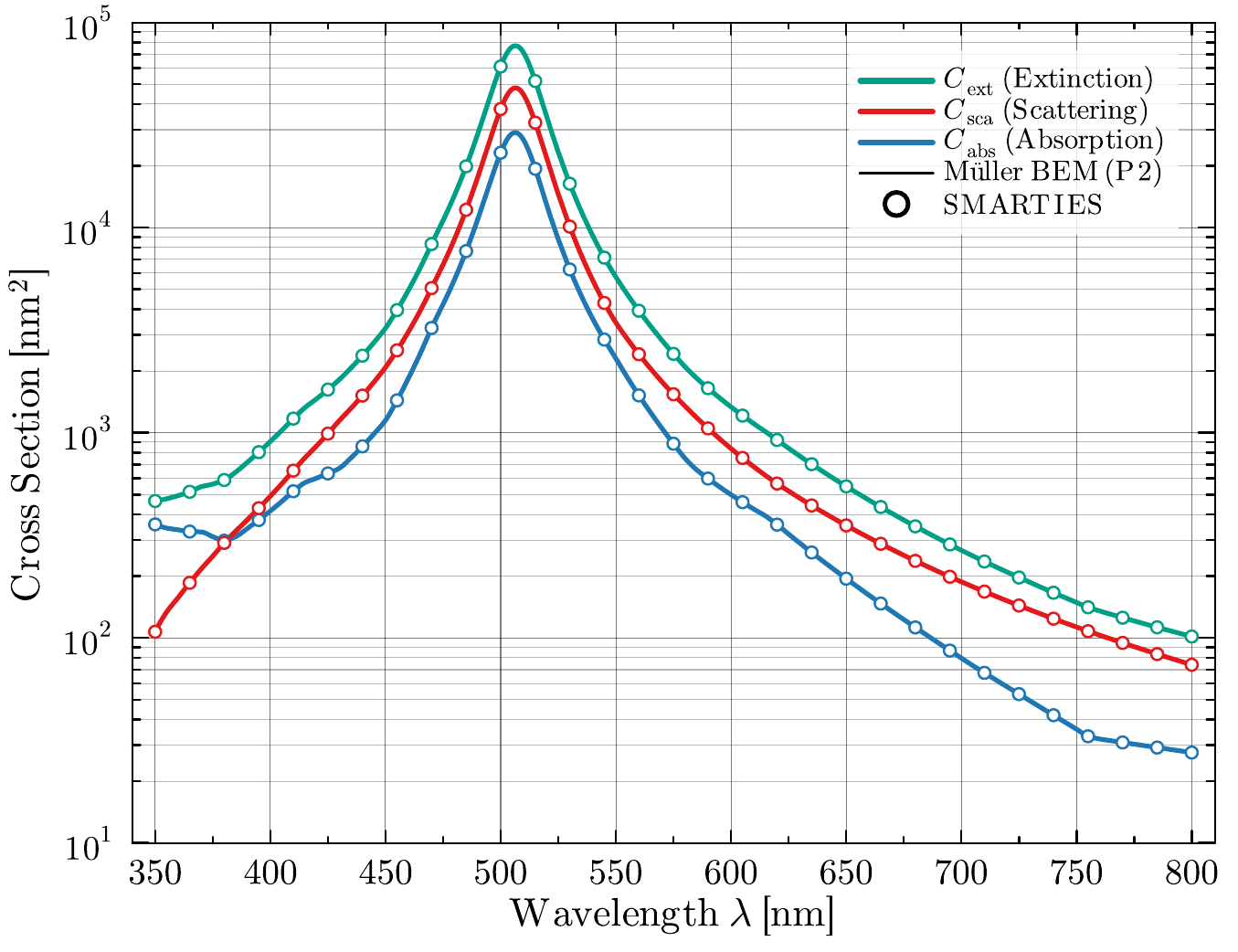}
    \caption{LSPR spectrum of the $30\times90$\,nm silver
             prolate spheroid. The cross sections $C_{\mathrm{ext}}$, 
             $C_{\mathrm{sca}}$, $C_{\mathrm{abs}}$ computed by the 
             M\"uller BIE (solid lines) are compared against SMARTIES 
             (open circles).}
    \label{fig:lspr_spectrum}
\end{figure}

Fig.~\ref{fig:lspr_spectrum} shows the extinction, scattering, and
absorption cross sections over the 350--800\,nm range. A wavelength
step of 1\,nm is used near the resonance (475--525\,nm) and 5\,nm
elsewhere. The M\"uller BIE results agree with the SMARTIES reference
across four decades of cross-section magnitude, reproducing the peak
extinction of ${\approx}80{,}000\,\mathrm{nm}^2$ at
$\lambda = 506\,\mathrm{nm}$. Away from the resonance, the cross
sections drop by nearly four orders of magnitude; the agreement is
maintained throughout. The optical theorem
$C_{\mathrm{ext}} = C_{\mathrm{abs}} + C_{\mathrm{sca}}$ is satisfied
at every wavelength.

\subsubsection{Preconditioning at the Resonance Peak}

\begin{figure}[htbp]
    \centering
    \includegraphics[width=0.7\columnwidth]{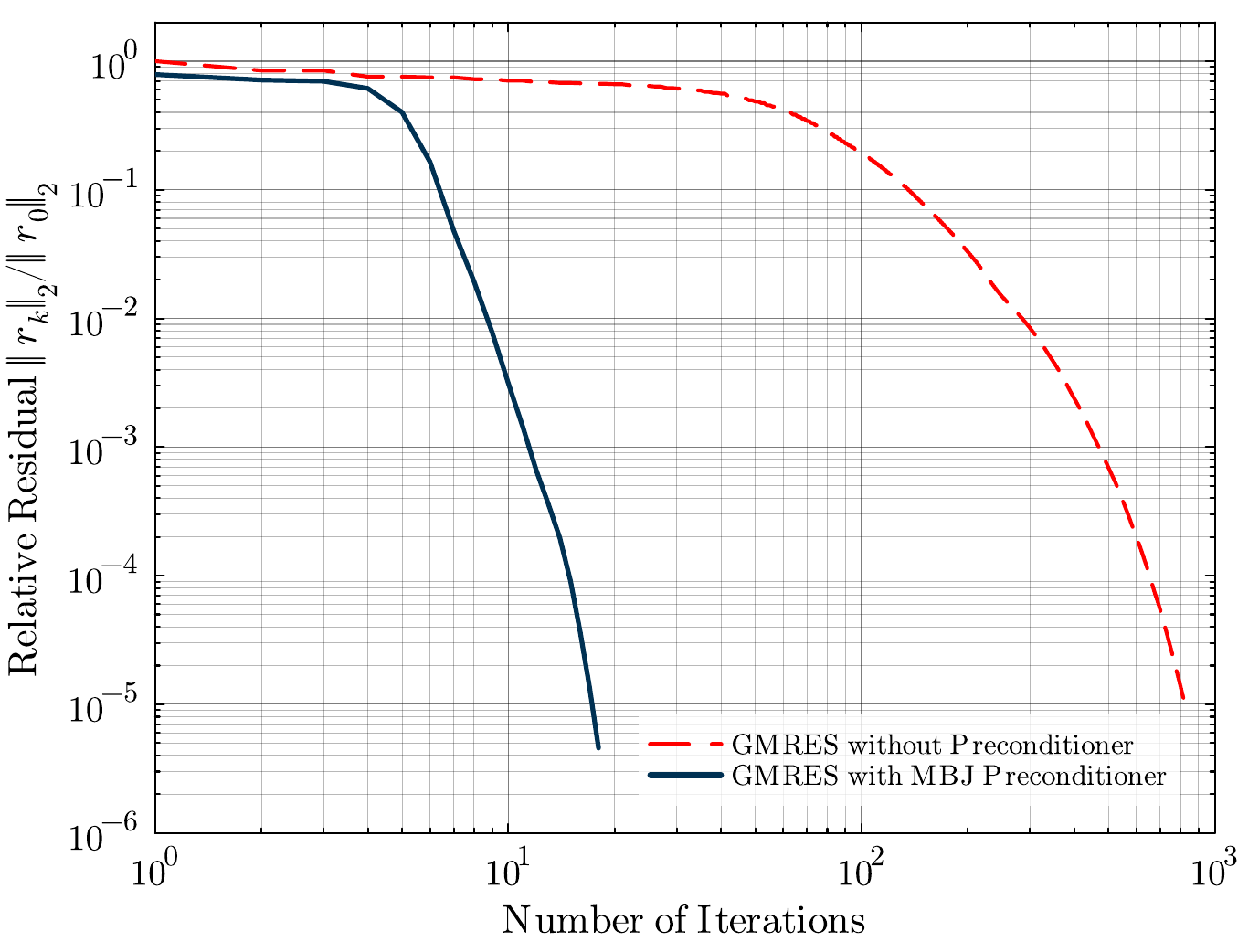}
    \caption{GMRES convergence at the 506\,nm LSPR peak.
             Dashed: unpreconditioned (822 iterations).
             Solid: MBJ preconditioned (18 iterations).}
    \label{fig:lspr_convergence}
\end{figure}

Near the LSPR peak at $\lambda = 506\,\mathrm{nm}$, the permittivity
reaches $\varepsilon_r \approx -10.15 + 0.32i$. The small imaginary
part relative to the large negative real part pushes the system toward
a physical resonance, deteriorating the spectral conditioning of the
discrete operator as eigenvalues approach the origin.
Unpreconditioned GMRES requires 822 iterations to reach a relative
residual of $10^{-5}$ (Fig.~\ref{fig:lspr_convergence}).

The MBJ preconditioner described in
Section~\ref{sec:preconditioning} is applied with a cluster size of
100 nodes, yielding 123 local blocks for the 24\,584-DOF system.
Their assembly and LU factorization require 0.37\,s and 75\,MB. The
preconditioned system converges in 18 iterations, a $45\times$
reduction, indicating that the MBJ preconditioner is effective
under the resonant spectral broadening induced by the extreme
permittivity.

\subsection{Benchmark~III: Elevated Electrical Size and Non-Convex Geometry}
\label{sec:val_extreme}

This benchmark examines two configurations that degrade the spectral conditioning of the discrete system. In Case~A, a convex dielectric at elevated electrical size is
considered~\cite{Nedelec2001, Ganesh2008}. The large ratio of
scatterer dimension to wavelength induces a highly oscillatory integral
kernel, increasing the modal density that the discrete operator must
resolve and spreading its eigenvalues in the complex plane. In Case~B,
a non-convex geometry with concavities is considered~\cite{ChandlerWilde2012, Betcke2014}. Surface regions that
are distant along the manifold geodesic are brought into close physical proximity, amplifying near-field interactions across the cavity and generating large-magnitude off-diagonal blocks that disrupt the diagonal dominance of the system. Both effects broaden the spectrum of the
discrete operator and inflate the required Krylov subspace dimension.

\subsubsection{Case~A: Dielectric Spheroid at Elevated Electrical Size}

\begin{figure}[htbp]
    \centering
    \includegraphics[width=0.7\columnwidth]{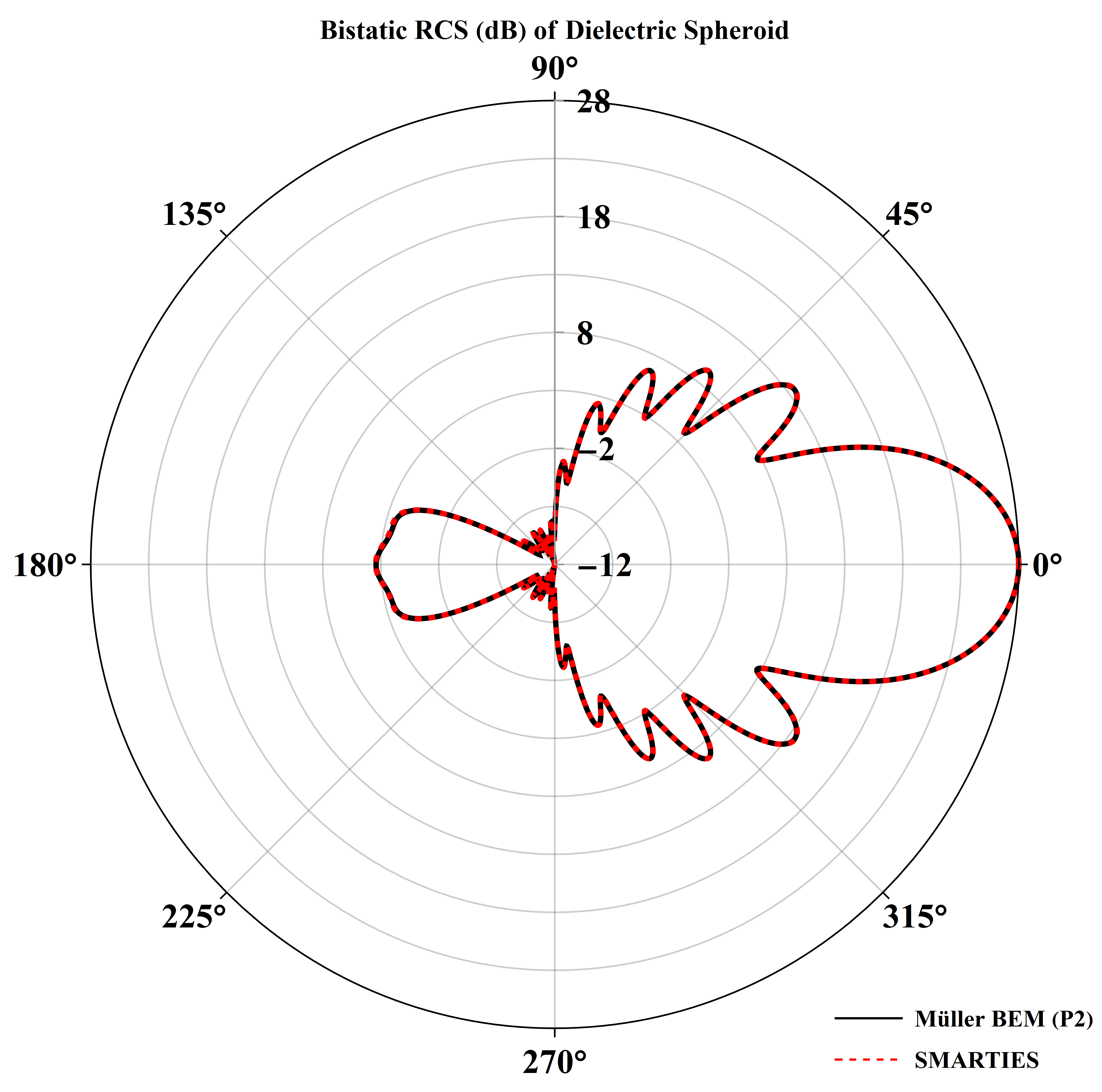}
    \caption{Bistatic RCS on $\varphi_s = 30^\circ$ for the 3:1
             dielectric spheroid ($k_a a = 6.0$, $k_a c = 18.0$).
             Solid: M\"uller BIE. Dashed: SMARTIES.}
    \label{fig:rcs_spheroid_ka6}
\end{figure}

A prolate spheroid with dimensionless semi-axes $a = b = 1$ and
$c = 3$ and lossless dielectric permittivity $\varepsilon_r = 2.25$
is considered at exterior size parameter $k_a a = 6.0$, corresponding
to a longitudinal electrical size of $k_a c = 18.0$. The interior
refractive index ($n = 1.5$) raises the internal electrical length to
$k_i c = 27.0$, producing a dense standing-wave structure within the
scatterer. The incident plane wave propagates along the major axis with Euler angles $(\theta_p, \phi_p, \alpha_p) = (0^\circ, 30^\circ, 60^\circ)$. The
surface is discretized into 4\,498 $\mathrm{P}_2$ elements, yielding
35\,992 DOFs. Fig.~\ref{fig:rcs_spheroid_ka6} compares the bistatic
RCS on the $\varphi_s = 30^\circ$ cut plane against the SMARTIES
reference. The M\"uller BIE accurately resolves the highly
oscillatory sidelobe structure over the full angular range.

Without preconditioning, GMRES requires 644 iterations (267.3\,s) to
reach the $10^{-5}$ tolerance. The MBJ preconditioner extracts 180
local blocks of 100 nodes (1.5\,s setup, 109.8\,MB), reducing the
iteration count to 36 (21.7\,s), an $18\times$ reduction in iterations
and $12\times$ acceleration in the solution phase.

\begin{figure}[htbp]
    \centering
    \includegraphics[width=0.7\columnwidth]{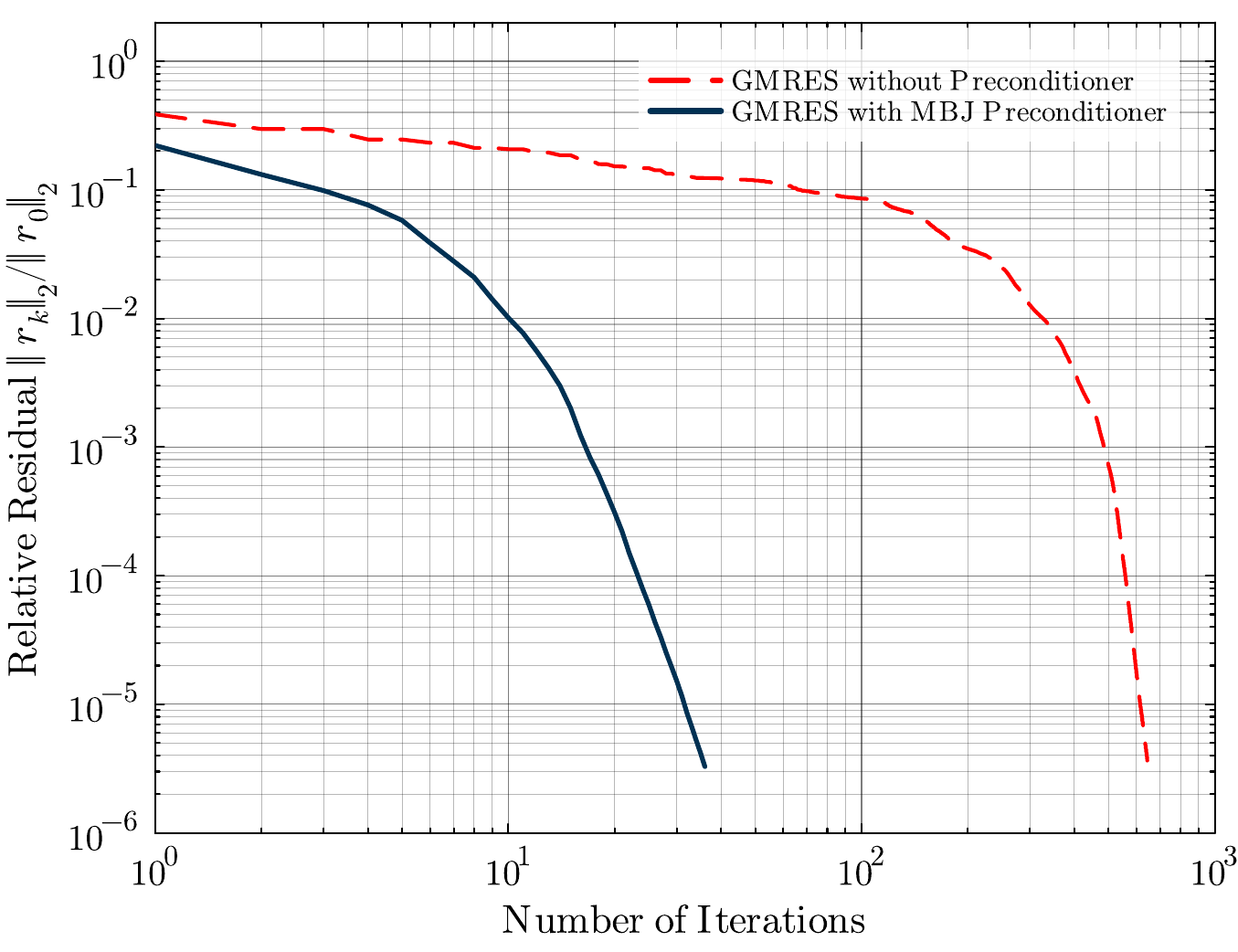}
    \caption{GMRES convergence for the 3:1 dielectric spheroid
             ($k_a c = 18.0$). Dashed: unpreconditioned (644
             iterations). Solid: MBJ preconditioned (36 iterations).}
    \label{fig:convergence_spheroid_ka6}
\end{figure}

\subsubsection{Case~B: Non-Convex Chebyshev Particle}

\begin{figure}[htbp]
    \centering
    \includegraphics[width=0.55\columnwidth]{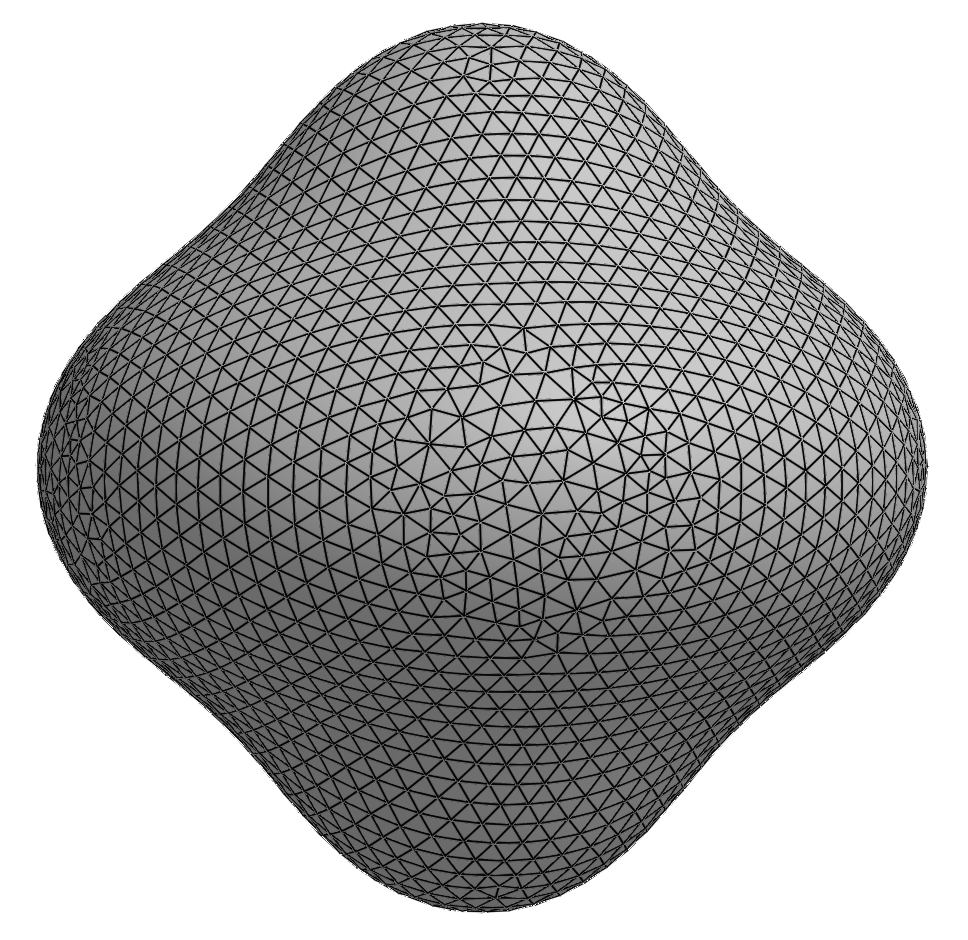}
    \caption{Isoparametric $\mathrm{P}_2$ mesh of the non-convex
             Chebyshev particle ($n{=}4$, $\eta{=}0.1$):
             10\,050 nodes, 5\,024 elements.}
    \label{fig:chebyshev_mesh}
\end{figure}

We next consider a geometry that combines elevated electrical size
with surface concavities. The Chebyshev particle of order $n$ is
defined by the radial profile~\cite{Wiscombe1986, Mishchenko1996,
Mishchenko2002}
\begin{equation}
r(\theta) = a\bigl(1 + \eta\cos(n\theta)\bigr),
\end{equation}
where $a$ is the unperturbed radius and $\eta$ the deformation
amplitude. With $n = 4$ and $\eta = 0.1$, the surface exhibits
four-fold polar lobes separated by concave valleys
(Fig.~\ref{fig:chebyshev_mesh}). The material is a lossless
dielectric ($\varepsilon_r = 2.25$) at exterior size parameter
$k_a a = 10.0$. The incident plane wave is parameterized by Euler
angles
$(\theta_p, \phi_p, \alpha_p) = (45^\circ, 30^\circ, 60^\circ)$. An
independent reference solution is computed using the EBCM, as
implemented in the open-source Julia package
\texttt{TransitionMatrices.jl}~\cite{TransitionMatrices2026}.

\begin{figure}[htbp]
    \centering
    \includegraphics[width=0.65\columnwidth]{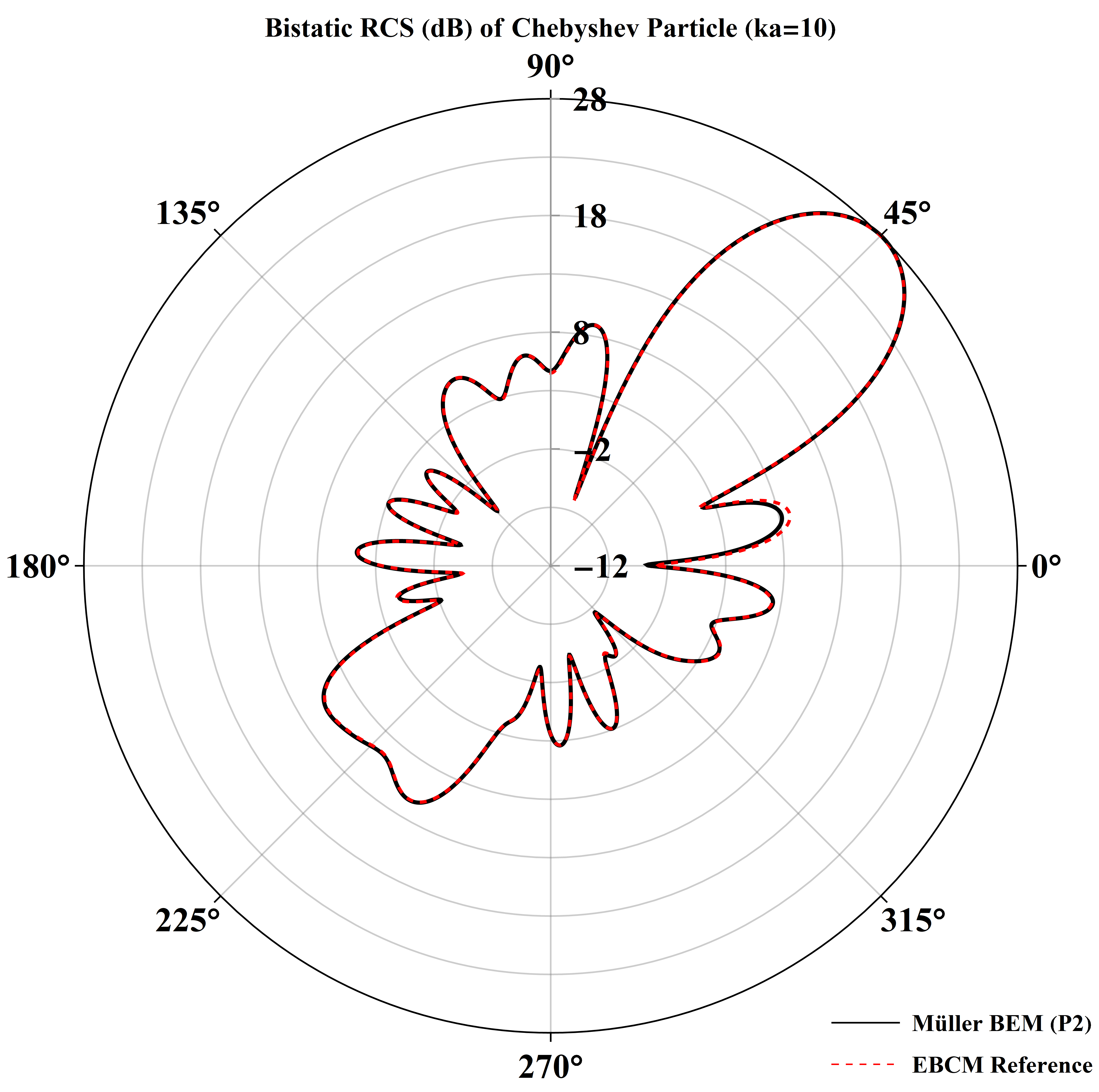}
    \caption{Bistatic RCS for the non-convex Chebyshev particle
             ($k_a a = 10.0$). Solid: M\"uller BIE. Dashed: EBCM.}
    \label{fig:rcs_extreme_chebyshev}
\end{figure}

Fig.~\ref{fig:rcs_extreme_chebyshev} compares the bistatic RCS on the
$\varphi_s = 30^\circ$ cut plane. The M\"uller BIE agrees with the
EBCM reference across the full dynamic range, reproducing both the
forward lobe and the fine oscillatory structure in the back hemisphere
arising from interactions across the non-convex lobes.

The 5\,024-element mesh yields 40\,200 DOFs, and the dense matrix
assembly completes in 102.4\,s. Unpreconditioned GMRES requires 950
iterations (474.9\,s) to converge. Applying the MBJ preconditioner
with 100-node clusters extracts 201 local blocks (2.3\,s setup,
122.7\,MB), reducing the iteration count to 71 and the solve time to
50.5\,s, a $13.4\times$ reduction in iterations and $9.4\times$
overall speedup.

\begin{figure}[htbp]
    \centering
    \includegraphics[width=0.7\columnwidth]{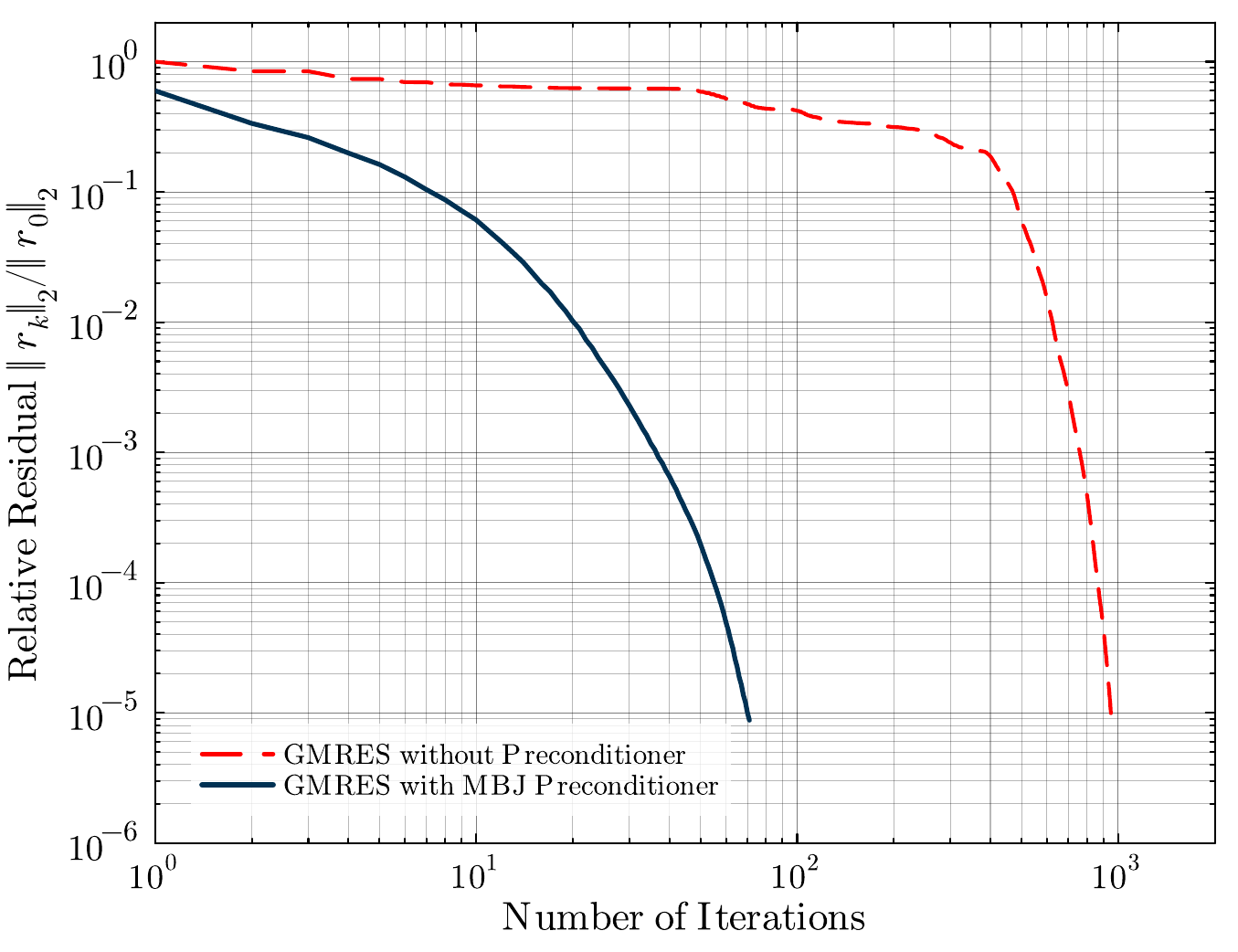}
    \caption{GMRES convergence for the non-convex Chebyshev particle.
             Dashed: unpreconditioned (950 iterations). Solid:
             MBJ preconditioned (71 iterations).}
    \label{fig:gmres_convergence}
\end{figure}

\section{Conclusion}
\label{sec:conclusion}

This paper has presented a nodal, high-order Galerkin boundary element
method for penetrable electromagnetic scattering. We demonstrated that
divergence-conforming basis functions are not required to discretize
the M\"uller formulation: because the Hessian acting on the kernel
difference $\varphi_a - \varphi_i$ has identical static limits, the
$\mathcal{O}(R^{-3})$ hypersingularity cancels at the operator level,
leaving only weakly singular $\mathcal{O}(R^{-1})$ remainders. This
allows a nodal discretization using $\mathrm{P}_2$ isoparametric
shape functions, with the surface vector field oriented by a
metric-weighted orthonormal tangent frame on curved manifolds. To
accelerate the iterative solution, a Morton-ordered Block Jacobi
preconditioner is applied. Each diagonal block preserves the full
$2 \times 2$ operator structure of the Galerkin system, rendering the
preconditioned operator a compact perturbation of the identity and
yielding superlinear GMRES convergence. Validation against
semi-analytical EBCM references confirms high-order spatial accuracy
and optical-theorem satisfaction to double-precision level under
varied material contrasts and geometries.

The present framework assumes smooth surfaces. Extending the
formulation to geometries with sharp edges and corners, where field
continuity fails and integrable singularities require dedicated
numerical treatment, is a natural direction for future work. A second
extension is the integration of the Fast Multipole Method (FMM) to
accelerate the far-field matrix-vector products and scale the solver
to larger problems.

\appendix
\refstepcounter{section}
\section*{Error Analysis of the Metric-Weighted Normal Estimation}
\label{app:covariant_normal}

\setcounter{equation}{0}
\setcounter{table}{0}
\renewcommand{\theequation}{\thesection\arabic{equation}}
\renewcommand{\thetable}{\thesection\arabic{table}}
\renewcommand{\thesubsection}{\thesection.\arabic{subsection}}

In Section~\ref{sec:manifold_basis}, a generalized metric weighting
scheme was introduced for nodal normal estimation. Classical
area-weighted normal averaging is susceptible to geometric noise on
distorted elements; the covariant approach avoids this instability.
This appendix uses differential geometry to prove that the metric
weight suppresses the divergence associated with extreme element
aspect ratios and, through a discrete Abel summation argument,
guarantees an $\mathcal{O}(h^2)$ convergence rate within a
shape-regular mesh family.

\subsection{Superconvergence of the Normal Direction on
$\mathrm{P}_2$ Manifolds}

Let $\Gamma$ be the exact smooth surface and let $v \in \Gamma$ be a
node with exact unit normal $\bm{n}$. Consider parametric curves
$\bm{\gamma}_k(s)$ emanating from $v$ along the element boundaries,
parameterized by arc length $s$ with $\bm{\gamma}_k(0) = \bm{x}_v$.
Each curve defines at $v$ a local Darboux frame
$(\bm{t}_k, \bm{u}_k, \bm{n})$, where
$\bm{t}_k = \bm{\gamma}_k'(0)$ is the unit tangent and
$\bm{u}_k = \bm{n} \times \bm{t}_k$ the tangent-normal. The second
derivative decomposes into normal and geodesic curvature components:
\begin{equation}
\bm{\gamma}_k''(0) = \kappa_{n,k}\,\bm{n} + \kappa_{g,k}\,\bm{u}_k.
\end{equation}
The Weingarten map in the Darboux frame~\cite{Lee2018} reads
$\bm{n}' = -\kappa_{n,k}\bm{t}_k - \tau_{g,k}\bm{u}_k$, so the
normal projection of the third derivative is
\begin{equation}
\bm{\gamma}_k'''(0) \cdot \bm{n}
  = \frac{d\kappa_{n,k}}{ds} + \kappa_{g,k}\,\tau_{g,k}
  \equiv C_k,
\end{equation}
where $\tau_{g,k}$ is the geodesic torsion. The quantity $C_k$ is a
local geometric invariant combining the normal curvature, geodesic
curvature, and geodesic torsion of the boundary curve at $v$.

For an isoparametric $\mathrm{P}_2$ element, the covariant tangent
vector $\bm{a}_k$ at the vertex is obtained from the parametric
derivative of the quadratic shape functions and is determined by the
vertex $\bm{\gamma}_k(0)$, the edge midpoint
$\bm{\gamma}_k(\ell_k/2)$, and the far node
$\bm{\gamma}_k(\ell_k)$, where $\ell_k \approx h$ is the metric
length of the edge:
\begin{equation}
\bm{a}_k = -3\bm{\gamma}_k(0) + 4\bm{\gamma}_k(\ell_k/2)
           - \bm{\gamma}_k(\ell_k).
\end{equation}
Substituting the Taylor expansion of $\bm{\gamma}_k(s)$ about
$s = 0$ yields
\begin{equation}
\bm{a}_k =\; \ell_k\,\bm{\gamma}_k'(0)
  + \left[4\!\left(\frac{\ell_k^2}{8}\right)
    - \frac{\ell_k^2}{2}\right]\bm{\gamma}_k''(0)
  + \left[4\!\left(\frac{\ell_k^3}{48}\right)
    - \frac{\ell_k^3}{6}\right]\bm{\gamma}_k'''(0)
  + \mathcal{O}(\ell_k^4).
\end{equation}
The quadratic coefficient vanishes exactly, independently of the
local curvature, through the $\mathrm{P}_2$ interpolation weights
$(-3, 4, -1)$. This eliminates the $\mathcal{O}(h^2)$ normal
curvature term and delays the out-of-plane component to
$\mathcal{O}(h^3)$:
\begin{equation}
\bm{a}_k \cdot \bm{n}
  = -\frac{\ell_k^3}{12}
    \bigl(\bm{\gamma}_k'''(0)\cdot\bm{n}\bigr)
    + \mathcal{O}(\ell_k^4)
  = -\frac{\ell_k^3}{12}\,C_k + \mathcal{O}(\ell_k^4).
\end{equation}
The $\mathrm{P}_2$ interpolant thus reproduces the surface tangent
plane to one order higher than expected, a superconvergence property.
Accordingly, the covariant vector decomposes into a principal tangent
component, an in-plane perturbation, and a cubic normal perturbation:
\begin{equation}
\bm{a}_k = \ell_k\,\bm{t}_k + \bm{e}_k^\parallel
           + \bm{e}_k^\perp,
\qquad
\bm{e}_k^\perp = -\frac{\ell_k^3}{12}\,C_k\,\bm{n}.
\label{eq:ak_decomposition}
\end{equation}

\subsection{Tangential Error and Stability Analysis}

The unnormalized surface normal at the $k$-th element vertex is
$\bm{A}_k = \bm{a}_k \times \bm{a}_{k+1}$. Substituting
\eqref{eq:ak_decomposition} generates several cross-product terms.
Because the in-plane perturbation $\bm{e}_k^\parallel$ lies in the
exact tangent space, its cross product with any principal tangent
vector is collinear with $\bm{n}$ and perturbs only the magnitude of
the facet normal, not its direction. The angular deviation is
therefore controlled by the out-of-plane perturbation
$\bm{e}_k^\perp$. Retaining the leading tangential terms gives
\begin{equation}
\bm{A}_k \approx \ell_k \ell_{k+1} \sin\theta_k\,\bm{n}
  + \ell_k(\bm{t}_k \times \bm{e}_{k+1}^\perp)
  - \ell_{k+1}(\bm{t}_{k+1} \times \bm{e}_k^\perp),
\end{equation}
where $\theta_k$ is the parametric vertex angle. The last two terms
lie in the tangent plane and define the tangential error
\begin{equation}
\bm{E}_k
  = \ell_k(\bm{t}_k \times \bm{e}_{k+1}^\perp)
  - \ell_{k+1}(\bm{t}_{k+1} \times \bm{e}_k^\perp).
\end{equation}
Substituting $\bm{e}_k^\perp = -(\ell_k^3 C_k / 12)\,\bm{n}$ and
using the Darboux tangent-normal
$\bm{u}_k = \bm{n} \times \bm{t}_k$ with the identity
$\bm{t}_k \times \bm{n} = -\bm{u}_k$ yields
\begin{equation}
\bm{E}_k
  = \frac{\ell_k \ell_{k+1}^3 C_{k+1}}{12}\,\bm{u}_k
    - \frac{\ell_{k+1} \ell_k^3 C_k}{12}\,\bm{u}_{k+1}.
\label{eq:tangential_error}
\end{equation}

Under standard area weighting, $\bm{E}_k$ is normalized by the facet
Jacobian $J_k \approx \ell_k \ell_{k+1} \sin\theta_k$, so the local
tangential error scales as $\mathcal{O}(h^2)/\sin\theta_k$. For
highly distorted elements ($\sin\theta_k \to 0$), this ratio
diverges, translating parametric skewness into unbounded geometric
noise in the estimated normal.

The metric weight circumvents this divergence. With
$w_k = (g_{11}^{(k)} g_{22}^{(k)})^{-1}
\approx (\ell_k^2 \ell_{k+1}^2)^{-1}$, the metric-weighted
tangential error $\tilde{\bm{E}}_k = w_k \bm{E}_k$ reduces to
\begin{equation}
\tilde{\bm{E}}_k
  = \frac{\ell_{k+1} C_{k+1}}{12\,\ell_k}\,\bm{u}_k
    - \frac{\ell_k C_k}{12\,\ell_{k+1}}\,\bm{u}_{k+1}.
\end{equation}
On a shape-regular vertex patch on a sufficiently smooth manifold,
the local geometric invariants ($C_k$, $\bm{u}_k$, $\theta_k$) are
uniformly bounded and the number of elements $M$ in the one-ring
neighborhood of $v$ is a finite constant independent of $h$. Summing
over the $M$ elements and applying a cyclic index shift
($M + 1 \equiv 1$) on the first term, a discrete Abel summation
extracts the common factor $\ell_k C_k / 12$:
\begin{equation}
\bm{E}_{\mathrm{tan}}
  = \sum_{k=1}^M \tilde{\bm{E}}_k
  = \sum_{k=1}^M \frac{\ell_k C_k}{12}
    \left(
      \frac{\bm{u}_{k-1}}{\ell_{k-1}}
        - \frac{\bm{u}_{k+1}}{\ell_{k+1}}
    \right).
\end{equation}

This representation reveals two properties. First, the
$(\sin\theta_k)^{-1}$ instability is eliminated: the normal estimate
cannot diverge regardless of local element skewness. Second, the
total tangential error is bounded. The prefactor $\ell_k C_k$ scales
as $\mathcal{O}(h)$, and the bracketed term is bounded by the
triangle inequality as
$\|\bm{u}_{k-1}\|/\ell_{k-1} + \|\bm{u}_{k+1}\|/\ell_{k+1}
= \mathcal{O}(h^{-1})$. With $M = \mathcal{O}(1)$, the accumulated
tangential error is
\begin{equation}
\|\bm{E}_{\mathrm{tan}}\|
  \sim \mathcal{O}(h) \times \mathcal{O}(h^{-1})
  = \mathcal{O}(1).
\end{equation}

Under bounded valence and non-degenerate facet angles
($\sin\theta_k \geq c > 0$), the principal normal component
accumulates as
\begin{equation}
\|\bm{N}_{\mathrm{normal}}\|
  \asymp \sum_{k=1}^M \frac{\sin\theta_k}{\ell_k \ell_{k+1}}
  = \mathcal{O}(h^{-2}),
\end{equation}
with negligible contributions from elements of large distortion or
large metric length. The angular deviation of the reconstructed
normal after normalization is therefore
\begin{equation}
\Delta\theta
  \approx \frac{\|\bm{E}_{\mathrm{tan}}\|}
               {\|\bm{N}_{\mathrm{normal}}\|}
  = \frac{\mathcal{O}(1)}{\mathcal{O}(h^{-2})}
  = \mathcal{O}(h^2).
\end{equation}
This establishes the $\mathcal{O}(h^2)$ convergence of the
metric-weighted normal estimation. By eliminating the
$(\sin\theta_k)^{-1}$ instability of area weighting, the scheme
remains robust against local element skewness within a shape-regular
mesh family.

For a spherical scatterer, $\kappa_n$ is constant and $\tau_g$
vanishes identically, so $C_k = \partial_s \kappa_n
+ \kappa_g \tau_g \equiv 0$. The leading tangential error vanishes,
and the convergence improves to $\mathcal{O}(h^3)$ without
algorithmic modification.

\bibliographystyle{unsrt} 
\bibliography{Refs}       

\end{document}